\documentclass[12pt]{article}
\usepackage{geometry}
\usepackage{setspace}
\usepackage{fancyhdr}
\usepackage{amsfonts}
\usepackage{amsmath}
\usepackage{verbatim}
\usepackage{listings}
\usepackage{lscape}
\usepackage{graphics}
\usepackage{tabularx}
\usepackage{float}
\usepackage{breqn}
\usepackage{wasysym}
\usepackage{fix-cm}
\usepackage{bbm}
\usepackage{color}
\usepackage{amsthm}
\usepackage{enumitem}
\usepackage{authblk}

\usepackage{appendix}

\usepackage{graphicx}

\usepackage{multirow}
\usepackage{parskip}
\usepackage{amsmath, amsthm, amssymb}
\usepackage{dsfont}
\usepackage{bm}
\usepackage{parskip}

\usepackage{framed,enumitem}
\usepackage{tikz}
\usetikzlibrary{positioning}
\usepackage{accents}
\usepackage{comment}
\usepackage[ruled,vlined]{algorithm2e}
\usepackage{breqn}

\usepackage[round]{natbib}

\newcommand{\abs}[1]{\left| #1 \right|}

\newtheorem{theorem}{Theorem}
\newtheorem{lemma}{Lemma}
\newtheorem{corollary}{Corollary}
\newtheorem{proposition}{Proposition}
\theoremstyle{definition}

\newtheorem{example}{Example}

\makeatletter
\newtheorem*{rep@theorem}{\rep@title}
\newcommand{\newreptheorem}[2]{%
\newenvironment{rep#1}[1]{%
 \def\rep@title{#2 \ref{##1}}%
 \begin{rep@theorem}}%
 {\end{rep@theorem}}}
\makeatother

\newreptheorem{theorem}{Theorem}
\newreptheorem{lemma}{Lemma}
\newreptheorem{corollary}{Corollary}
\newreptheorem{proposition}{Proposition}

\providecommand{\keywords}[1]
{
  \small	
  \textbf{\textit{Keywords---}} #1
}

\newcommand{\calS}{\mathcal{S}}

\title{Positive-definiteness in separable priors: effects on prior interpretability and inference}
\author[1,2,*]{Jack Storror Carter}
\author[1,2]{David Rossell}
\affil[1]{Dept. of Business and Economics, Universitat Pompeu Fabra, Barcelona, Spain}
\affil[2]{Data Science Center, Barcelona School of Economics, Barcelona, Spain}
\affil[*]{Correspondence to jack.carter@upf.edu}
\date{}

\begin{document}

\maketitle

\begin{abstract}

A popular class of priors for symmetric positive-definite matrices assumes independent entries and adds a truncation to ensure positive-definiteness.
While conceptually simple and often computationally convenient, unless done carefully this truncation can have unintended effects.
If the truncated prior or its margins are significantly different from their untruncated counterpart, then its interpretability may suffer, its shrinkage properties become harder to characterise, and posterior inference may be affected in unanticipated ways. We investigate the effect of the truncation both for dense and sparse matrices, and show how to set prior parameters such as the variance of off-diagonal entries such that said effect is mitigated as the matrix dimension grows. We pay particular attention to sparse inference where, unless prior parameters are set carefully, the truncated prior and hence its corresponding posterior assign systematically higher mass to sparser structures than the untruncated prior.


\end{abstract}


Distributions over the space of symmetric positive-definite matrices play an important role in Bayesian statistics. This is mostly for their use as priors for covariance matrices and their transformations, such as correlation matrices, precision matrices or partial correlation matrices, but positive-definite matrices also appear as parameters in many multivariate and matrix-variate distributions such as the Wishart, matrix t and matrix F distributions. 
Let $\Theta = (\theta_{ij})$ be a random $k \times k$ symmetric and positive-definite matrix.
A conceptually simple class of prior distributions assumes that the density is
\begin{align} 
p^+(\Theta) = \frac{p(\Theta) \mathbb{I}(\Theta \succ 0)}{c}
\label{eq:pd_separable_prior}
\end{align}
where $\mathbb{I}(\Theta \succ 0)$ is an indicator for $\Theta$ being positive-definite,
$$ p(\Theta) = \prod_{i \leq j} \pi_{ij}(\theta_{ij}),$$
and $\pi_{ii}: \mathbb R^+ \to \mathbb R^+$ 
and $\pi_{ij}: \mathbb R \to \mathbb R^+$ for $i<j$ are density functions.
In \eqref{eq:pd_separable_prior}, the normalising constant $c = \mathbb{P}_p(\Theta \succ 0)= \int \mathbb{I}(\Theta \succ 0) d P(\Theta)$ is the probability that $\Theta \succ 0$ under $P$, the distribution whose density is $p$.
That is, $p$ assumes that the entries of $\Theta$ are independent and $p^+$ truncates $p$ onto the space of positive-definite matrices.
We refer to $p$ and $p^+$ as \textit{separable} and \textit{positive-definite separable} (PD-separable) distributions respectively. 
We denote the marginal distribution of $\theta_{ij}$ under $p^+(\Theta)$ as $\pi^+_{ij}(\theta_{ij})$.
In the identical distributed case $\pi_{ii} = \pi_D$ for all $i$ and some $\pi_D$, and $\pi_{ij} = \pi$ for all $i < j$ and some common off-diagonal $\pi$, then also
$\pi^+_{ii} = \pi^+_D$ for all $i$ and $\pi^+_{ij} = \pi^+$ for all $i < j$, for some common $(\pi_D^+, \pi^+)$. 

A setting in which PD-separable priors generated particular interest are graphical models.
For example, in Gaussian graphical models with precision (inverse covariance) matrix $\Theta$, additionally to estimating $\Theta$, one also seeks to detect the zero off-diagonal entries in $\Theta$ (its sparsity structure).  
The interest in PD-separable priors stems partly
from their perceived interpretability, and perharps more importantly for their leading to computationally-convenient posterior sampling algorithms and posterior mode estimators that are directly connected to penalised likelihood methods.
For example, one may obtain the posterior mode for the Bayesian graphical LASSO of \cite{Wang2012} using standard graphical LASSO \citep{Friedman2008} algorithms, or alternatively use algorithms that encourage the presence of positive partial correlations \citep{lauritzen:2020}.
\cite{Khondker2013} extended the work of \cite{Wang2012}, and \cite{Banerjee2015,Wang2015,Gan2018,sulem2025bayesian} considered various PD-separable spike-and-slab priors, along with computational proposals.
Beyond canonical Gaussian graphical models, PD-separable priors also found use for correlation matrices \citep{azose2018estimating},
for regressing graphical models on network data \citep{Jewson2022}
and for graphical models for mixed data types such as combinations of continuous, categorical and count variables \citep{florez:2025}.

In this paper we argue that, while the construction of $p^+$ is appealingly simple, its interpretation is not.
Beyond the fact that the entries of $\Theta$ are no longer independent, as intended to ensure that $\Theta$ is positive-definite, there may be unintended consequences.
Specifically, the marginal distribution $\pi_{ij}^+$ can be dramatically different from $\pi_{ij}$, and the prior beliefs about sparsity structures in $\Theta$ may be significantly affected, with the consequent effects on posterior inference. 
For example, \cite{Wang2012} showed that when $\pi_{ii}$ are exponential and $\pi_{ij}$ Laplace distributions, the truncation can significantly inflate the diagonal of $\Theta$ (relative to $\pi_{ii}$) and shrink its off-diagonal entries towards 0 (relative to $\pi_{ij}$).
This has practical implications.
First, an important appeal of the Bayesian paradigm is that prior hyper-parameters often describe interpretable characteristics of the data-generating process, such as the proportion of non-zero entries in $\Theta$, or their variance.
In contrast to approaches where hyper-parameters are treated in a black-box manner, one can use the hyper-parameters to incorporate prior knowledge.
Further, one can describe quantities of interest (e.g., the proportion of non-zeroes in $\Theta$) by obtaining point estimates for the hyper-parameters (or their full posterior distribution).
Second, the fact that the truncation can significantly alter the prior implies that the prior shrinkage may have unanticipated effects, or simply be much stronger or milder than one may naively expect, which has an effect on posterior inference.
As we show, this is particularly obvious for sparse structural learning tasks.

Our first main contribution is describing precisely how $p^+$ and its margins $\pi_{ij}^+$ differ from the untruncated $p$ and $\pi_{ij}$, and providing a series of results that show how to set prior parameters such that these differences are small.
From \eqref{eq:pd_separable_prior}, if $c=1$ then $p=p^+$.
We show that the key aspects for ensuring that $c$ is close to 1 are having sufficiently small off-diagonal variances or a sufficiently large minimum diagonal entry, and give conditions ensuring that $c \to 1$ as the matrix dimension $k \to \infty$.
Our second main contribution is describing the impact of the positive-definite truncation in sparse settings. More specifically, on the prior distribution of the sparsity pattern (zero entries) of $\Theta$, a key feature for structural learning in graphical models.
We show that sparsity in $\Theta$ generally increases the probability of positive-definiteness, and the immediate implications on the prior probability of sparsity patterns under $p^+$ relative to under $p$.
We also show that setting a large off-diagonal variance, in the limit, results in assigning posterior probability 1 to $\Theta$ being a diagonal matrix.
We obtain bounds for the probability of $\Theta \succ 0$ that hold for any fixed $k$ and sparsity structure, which indicate a possible way to set the prior off-diagonal variance to avoid such pathologies. 
Let $\mu$ be the mean of the diagonal $\pi_D$, $\sigma$ the variance of off-diagonal $\pi$ and $d_z$ the maximum row degree of $\Theta$ (defined below). Roughly speaking, whereas the dense case requires $\sigma / \mu < 1/(2\sqrt{k})$ for $c$ to converge to 1, in the sparse case it suffices that $\sigma/\mu = o(1/\sqrt{d_z})$ if $\Theta$ has fixed diagonal and $\sigma/\mu= o(1/(k \sqrt{d_z})$ if $\theta_{ii} \sim \mbox{Exp}(1/\mu)$ (neglecting log terms).




The paper proceeds as follows.
In Section \ref{sec:charact} we describe how $p^+$ relates to $p$. We explain that the total variation distance (TVD) between $p^+$ and $p$ and their Kullback-Leibler (KL) divergence are simple functions of $c$.
We then give a simple expression relating the marginal $\pi_{ij}^+$ to the untruncated $\pi_{ij}$, and show that their TV distance and KL divergence are upper-bounded by functions of $c$.
These results make explicit precisely how, as $c$ approaches 1, $p^+$ converges to $p$, but also how in some settings the marginals $\pi_{ij}$ and $\pi_{ij}^+$ can remain close even when $c$ approaches 0.
In Section \ref{sec:contolc} we describe how to set prior parameters such that $c$ is close to 1, in particular showing that the key is to set a sufficiently small prior variance on off-diagonal $\theta_{ij}$.
The results in Section \ref{sec:charact}-\ref{sec:contolc} hold for any prior of the form in \eqref{eq:pd_separable_prior}.
Section \ref{sec:sparse_matrices} considers the important case of sparse matrices, where $\theta_{ij}=0$ has positive prior probability.
We describe the effect of the truncation on posterior inference, specifically via its effect on the prior mass assigned to each sparsity structure in $\Theta$, which it turns depends on the prior probability that $\Theta \succ 0$ under that structure.
We show that, for any given sparsity structure of $\Theta$, said probability of $\Theta \succ 0$ is a decreasing function of the off-diagonal prior variance.
We also show that said probability decreases as one adds non-zero entries in $\Theta$, give specific conditions for the probability to converge to 0 or 1,
and obtain lower and upper bounds that hold for any sparsity pattern and dimension $k$.
All proofs are in the appendix.

{\bf Notation.}
We denote the smallest eigenvalue of $\Theta$ by $\lambda_{\min}(\Theta)$,
and let $\theta_{\min}=\min_{i=1,\ldots,k} \theta_{ii}$ and $\theta_{\max}= \max_{i=1,\ldots,k} \theta_{ii}$.
We define the binary $k \times k$ matrix $Z=(z_{ij})$ with zero diagonal and $z_{ij}= \mathbb I(\theta_{ij} \neq 0)$ for $i \neq j$,
and denote the degree (or number of non-zero off-diagonal entries) of row $i$ by $d_i = \sum_{j \neq i} z_{ij}$,
and the maximum edge degree by $d_z= \max_{i \in \{1,\ldots,k\}} d_i$.
We denote by $\mathbb P_p(A)$ the probability of event $A$ under the distribution or density $p$, and by $\mathbb E_p[]$ the corresponding expectation.
$\delta_0$ denotes a Dirac measure at 0.
For two positive sequences $a_k, b_k$, we denote by $a_k= o(b_k)$ when $\lim_{k \to \infty} a_k/b_k = 0$.

\section{Characterising $p^+$}\label{sec:charact}



The normalising constant $c \in [0,1]$ in \eqref{eq:pd_separable_prior} is an important quantity because it determines both the total variation (TV) distance and Kullback-Leibler (KL) divergence between $p^+$ and $p$. This is a standard result for all truncated distributions, but we re-state and prove it here for our specific case.
\begin{proposition}\label{prop:dist}
    The TV distance and KL divergence between $p$ and $p^+$ are
    \begin{align*}
        \mbox{TV}(p^+,p) = 1 - c, && KL(\,p^+ \;||\; p\,) = - \log c,
    \end{align*}
    and $p^+$ minimises both distances among all distributions on the set of positive-definite matrices. 
\end{proposition}

Proposition \ref{prop:dist} makes explicit how, if $c$ is close to 1, then $p^+$ is close to $p$.
In contrast, if $c$ is close to 0 then $p^+$ may encode very different prior beliefs and shinkage properties than $p$.
Importantly, it is possible for the marginal $\pi_{ij}^+$ to remain close to $\pi_{ij}$, even when the joint $p^+$ is not close to $p$. 
In such a case, one may argue that the beliefs and shrinkage encoded by $p$ and $p^+$ are not that different, in that they differ mainly in their dependence structure (as one expects, since the $\Theta \succ 0$ constraint induces dependence) but less so in their marginals.
The remainder of this section studies the behavior of these marginals.


First, Proposition \ref{prop:marg} relates
$\pi^+_{ij}(\theta_{ij})$ to the untruncated $\pi_{ij}(\theta_{ij})$ using 
$c_{ij}(\theta_{ij}) = \mathbb{P}_p\left(\, \Theta \succ 0 \mid \theta_{ij} \,\right)$.
Specifically, the ratio $\pi_{ij}^+(\theta_{ij})/\pi_{ij}(\theta_{ij}) = c_{ij}(\theta_{ij})/c$.
\begin{proposition}\label{prop:marg}
    The marginal density of $\theta_{ij}$ under $p^+$ is
     $$ \pi^+_{ij}(\theta_{ij}) = \frac{c_{ij}(\theta_{ij})}{c}\, \pi_{ij}(\theta_{ij}).$$
     Further, $\mathbb{E}_{\pi_{ij}}[c_{ij}(\theta_{ij})] = c$,
each diagonal $c_{ii}(\theta_{ii})$ is non-decreasing in $\theta_{ii}$
and $\lim_{\theta_{ii} \to 0} c_{ii}(\theta_{ii})=0$.
\end{proposition}



Proposition \ref{prop:marg} also shows that $c_{ij}(\theta_{ij})/c$ is equal to 1 in expectation under $\pi_{ij}$, as expected since $\pi_{ij}^+$ assigns higher density to some values of $\theta_{ij}$ than the untruncated $\pi_{ij}$, and hence lower density to other values.
Interestingly, Proposition \ref{prop:margindist} shows that 
the TV distance and KL divergence between $\pi_{ij}^+$ and $\pi_{ij}$
depend on the distribution of $c_{ij}(\theta_{ij})/c$ in a very specific and simple manner.
\begin{proposition}\label{prop:margindist}
    The TV distance and KL divergence between $\pi_{ij}$ and $\pi^+_{ij}$ are
    \begin{align*}
        \mbox{TV}(\pi_{ij}^+, \pi_{ij}) = \frac{1}{2} \mathbb{E}_{\pi_{ij}}\abs{ \frac{c_{ij}(\theta_{ij})}{c} - 1} && KL(\,\pi_{ij}^+ \;||\; \pi_{ij}\,) = \mathbb{E}_{\pi_{ij}} \left[ \frac{c_{ij}(\theta_{ij})}{c} \log\left( \frac{c_{ij}(\theta_{ij})}{c} \right) \right]
    \end{align*}
\end{proposition}
Despite its simplicity, Proposition \ref{prop:margindist} involves the distribution of $c_{ij}(\theta_{ij})/c$, which is not available in closed-form.
Simpler trivial upper-bounds are available, using that the TV distance and KL divergence of a marginal is upper-bounded by that of the joint.
\begin{align*}
    \mbox{TV}(\pi_{ij}^+, \pi_{ij}) \leq 1 - c && KL(\,\pi_{ij}^+ \;||\; \pi_{ij}\,) \leq -\log c.
\end{align*}
Hence a sufficient condition for the marginals to be close is that $c \approx 1$. 
However, Propositions \ref{prop:marg} and \ref{prop:margindist} show that the margins can remain close for smaller $c$, provided that $c_{ij}(\theta_{ij}) \approx c$ for all $\theta_{ij}$.

Proposition \ref{prop:margindist_upperbound} provides much sharper bounds, particularly in the identically-distributed case. It first bounds the mean marginal distance by a function of $c$. It shows that, even if one has vanishing $c$ as $k$ grows (equivalently, $-\log c \to \infty$), and hence $p^+$ becomes very different from $p$, as long as $- \log c= o(k^2)$ we have that the mean marginal distance tends to 0. However, without further structure this does not prohibit an individual marginal distance from being arbitrarily large. Assuming identically distributed entries does provide this structure, resulting in all off-diagonal distances vanishing if $- \log c= o(k^2)$ and all diagonal distances vanishing if $- \log c= o(k)$.


\begin{proposition}\label{prop:margindist_upperbound}
The mean marginal KL divergence and TV distance are upper-bounded by
\begin{align*}
    &\frac{2}{k(k+1)}\sum_{i \leq j} KL(\,\pi_{ij}^+ \;||\; \pi_{ij}\,) \leq \frac{-2 \log c}{k(k+1)}; 
    && \frac{2}{k(k+1)}\sum_{i \leq j} \mbox{TV}(\pi_{ij}^+, \pi_{ij}) \leq \min \left\{ \sqrt{\frac{-\log c}{k(k+1)}}, \sqrt{1 - c^{2/k(k+1)} } \right\}.
\end{align*}
In the identically-distributed case where $\pi_{ii}=\pi_D$ and $\pi_{ij}=\pi$ for $i<j$, we have that
\begin{align*}
    &KL(\,\pi_{D}^+ \;||\; \pi_{D}\,) \leq \frac{-\log c}{k} && KL(\,\pi^+ \;||\; \pi\,) \leq \frac{-2 \log c}{k(k-1)}
\\
    &\mbox{TV}( \pi_D^+, \pi_D) \leq \min\left\{ \sqrt{\frac{-\log c}{2k}}, \sqrt{ 1 - c^{1/k} } \right\} && 
    \mbox{TV}( \pi^+, \pi) \leq \min\left\{ \sqrt{\frac{-\log c}{k(k-1)}}, \sqrt{ 1 - c^{2/k(k-1)} } \right\}.
\end{align*}
\end{proposition}

The bounds on the TV distances involve two terms.
The first term is sharper when $c$ is close to 1, that is when the distances are small, while the second is sharper for large distances ($c$ is close to 0).

\section{Controlling $c$}\label{sec:contolc}

In Section \ref{sec:charact} we saw that 
$c = \mathbb{P}_p(\Theta \succ 0)$ determines the distance between $p$ and $p^+$, and that it upper-bounds the distance between the marginals $\pi_{ij}$ and $\pi_{ij}^+$. 
Although not denoted explicitly for simplicity, $c$ is a function of $k$ and of $p$.
For a given $k$ and $p$, the value of $c$ can be numerically approximated by sampling from $p$.
Here we obtain a richer description by obtaining bounds on $c$, portraying its behavior as $k \to \infty$ and as one varies characteristics of $p$ such as the mean of the diagonal $\pi_{ii}$ or the variance of the off-diagonal $\pi_{ij}$. 

First, we state Propositions \ref{prop:cbounds}-\ref{prop:sigma}, which give elementary bounds showing that $c$ grows as the off-diagonal variance decreases, or as the diagonal entries shift towards larger values.
We consider general distributions and then specialize our results for the case where the off-diagonal $\pi_{ij}$ follow Gaussian (Proposition \ref{prop:cbounds}) and more general scale families (Proposition \ref{prop:sigma}). 
Recall that a random variable $W$ follows a scale family with parameter $\sigma$ and base distribution $f$ if it is equal in distribution to $\sigma W_0$, where $W_0$ is $f$-distributed.
Second, we state Theorems \ref{thm:Wigner}-\ref{thm:Wigner2}, which outline the conditions under which $c$ converges to either 0 or 1 as $k \to \infty$, for an ample family of distributions.
Altogether, our results give explicit expressions on how to vary the diagonal $\pi_{ii}$ and off-diagonal $\pi_{ij}$ such that $c$ is close to 1, and that therefore one can easily interpret the prior beliefs and prior shrinkage associated to $p^+$.

A sufficient condition for $\Theta \succ 0$ is that the diagonal dominates the off-diagonal, i.e. $\theta_{ii} > \sum_{j \neq i} \abs{\theta_{ij}}$ for each $i$. This condition is guaranteed to hold if each $\abs{\theta_{ij}} < \theta_{\min}/(k-1)$ where $\theta_{\min} = \min_{i=1,\dots,k} \theta_{ii}$. Hence 
\begin{align*}
    c &\geq \mathbb{P}_p\left( \bigcap_{i=1,\dots,k} \left\{ \theta_{ii} > \sum_{j \neq i} \abs{\theta_{ij}} \right\} \right) 
    \geq \mathbb{P}_p\left( \bigcap_{i < j} \left\{ \abs{\theta_{ij}} < \theta_{\min} / (k-1) \right\} \right).
\end{align*}
On the other hand, a necessary condition for $\Theta \succ 0$ is that $\abs{\theta_{ij}} < \sqrt{\theta_{ii}\theta_{jj}}$ for all $i < j$. 
Therefore, 
\begin{align*}
    c &\leq \mathbb{P}_p\left( \bigcap_{i < j} \left\{ \abs{\theta_{ij}} < \sqrt{\theta_{ii}\theta_{jj}} \right\} \right) 
    \leq \mathbb{P}_p\left( \bigcap_{i<j} \left\{ \abs{\theta_{ij}} < \theta_{\max} \right\} \right)
\end{align*}
where $\theta_{\max} = \max_{i=1,\dots,k}\theta_{ii}$.
Proposition \ref{prop:cbounds} states these bounds for the Gaussian off-diagonal $\pi_{ij}$ case.

\begin{proposition}\label{prop:cbounds}
    Let $p(\Theta)$ be a separable distribution with identically distributed off-diagonals $\pi(\theta_{ij}) = N(\theta_{ij} ; 0, \sigma^2)$. 
    Then
    \begin{align*}
        \mathbb{E}_p \left[ \left( 2 \Phi \left( \theta_{\min}/(k-1) \sigma \right) - 1 \right)^{k(k-1)/2} \right] \leq c \leq \mathbb{E}_p \left[\left( 2 \Phi \left( \theta_{\max}/\sigma \right) - 1 \right)^{k(k-1)/2}\right],
    \end{align*}
    where the expectations are taken with respect to $\theta_{\min}$ and $\theta_{\max}$ respectively.
\end{proposition}

Proposition \ref{prop:cbounds} shows that 
$c$ can be controlled by the off-diagonal variance $\sigma^2$
and the distribution of the minimum diagonal entry. 
Specifically, one can ensure that $c$ is close to 1 by setting small $\sigma^2$ and ensuring that the $\pi_{ii}$'s place sufficient mass away from 0.
An important particular case is when $\Theta$ has fixed diagonal entries $\theta_{ii}=1$, for example when $\Theta$ is a correlation or partial correlation matrix. Then
\begin{align*}
    \left( 2 \Phi \left( 1/(k-1) \sigma \right) - 1 \right)^{k(k-1)/2} 
    \leq c \leq 
    \left( 2 \Phi \left( 1/\sigma \right) - 1 \right)^{k(k-1)/2}.
\end{align*}
Note that for fixed $\sigma$, the upper bound implies that $-\log c \neq o(k^2)$. Hence, for the bounds in Proposition \ref{prop:margindist_upperbound} to imply that $\mbox{TV}(\pi,\pi^+) \to 0$ requires that $\sigma \to 0$.

When the density functions of $\theta_{\min}$ and $\theta_{\max}$ have closed form, the expectations can be written in terms of the corresponding integrals. For example, for the choice $\pi_D(\theta_{ii}) = \mathrm{Exp}(\theta_{ii};\lambda)$ used by \cite{Wang2015,sulem2025bayesian}, we have
\begin{align*}
        k\lambda \int_0^\infty \left( 2 \Phi \left( x/(k-1) \sigma \right) - 1 \right)^{k(k-1)/2} e^{-k\lambda x} \, dx 
        \leq c \leq 
        k \lambda\int_0^\infty \left( 2 \Phi \left( x/\sigma \right) - 1 \right)^{k(k-1)/2} e^{-\lambda x}(1-e^{-\lambda x})^{k-1} \, dx,
\end{align*}

These results also show that, if $\sigma^2$ and the distribution of $\theta_\min$ are fixed as $k \to \infty$, then $\lim_{k \to \infty} c= 0$. 
Proposition \ref{prop:sigma} extends this result 
to the more general scale family. 
\begin{proposition}\label{prop:sigma}
    Suppose that $\pi_{ij}(\theta_{ij})$ for each $i<j$ is a continuous distribution within the scale family with zero mean and common scale parameter $\sigma$ (shared across $i<j$).
    Then for any fixed diagonal distributions $\pi_{ii}$, $c$ is continuous and strictly decreasing in $\sigma$. Further, $\lim_{\sigma \to 0} c= 1$ and $\lim_{\sigma \to \infty} c= 0$.
\end{proposition}

To give more precise conditions on how $\sigma$ affects $c$, we turn to the theory of Wigner matrices, a class of random matrices for which, under mild assumptions, the eigenvalues converge to a fixed semicircle distribution as the matrix dimension $k \to \infty$. 
Consider first the case where $\Theta$ has fixed diagonal $\theta_{ii}=\mu > 0$. Similar to $\sigma$, $\mu$ may depend on the dimension $k$ but we omit this dependence from the notation for simplicity.
Suppose that the off-diagonal entries are $\theta_{ij}= \sigma \tilde{\theta}_{ij}$, where $\tilde{\theta}_{ij} \sim \tilde{\pi}$, a distribution with zero mean and unit variance and which does not depend on $k$.
Theorem \ref{thm:Wigner} shows how the limiting behavior of
$c = \mathbb{P}_p(\Theta \succ 0)$ as $k$ grows depends on $(\mu,\sigma,k)$.
Its proof relies on Wigner matrix theory, which gives limiting results for the eigenvalues of $\widetilde{\Theta}=(\tilde{\theta}_{ij})$ where $\tilde{\theta}_{ii}=0$, and therefore for those of $\Theta= \mu I + \sigma \widetilde{\Theta}$.

\begin{theorem}\label{thm:Wigner}
Let $\Theta= \mu I + \sigma \widetilde{\Theta}$, where $\mu>0$, $\sigma \geq 0$, $\tilde{\theta}_{ii}=0$ and $\tilde{\theta}_{ij} \sim \tilde{\pi}$ independently for $1 \leq i < j < \infty$, where $\tilde{\pi}$ has mean zero, variance one, finite fourth moment and does not depend on $k$.
\begin{enumerate}[label=\roman*)]
    \item If $\liminf_{k \rightarrow \infty} \frac{ \mu }{ \sigma \sqrt{ k } } > 2$ then $\lim_{k \to \infty} c = 1$.
    \item If $\limsup_{k \rightarrow \infty} \frac{ \mu }{ \sigma \sqrt{ k } } < 2$ then $\lim_{k \to \infty} c = 0$.
\end{enumerate}

Further, if one sets any $\sigma = \frac{\mu}{(2+\delta_k) \sqrt{k}}$ where $\delta_k$ satisfies $k^{-2/3}= o(\delta_k)$, then $\lim_{k \to \infty} c= 1$.
\end{theorem}

Writing $\frac{ \mu }{ \sigma \sqrt{ k } } = 2 + \delta_k$, Theorem \ref{thm:Wigner} shows that the limit 
of $c$ depends on whether $\delta_k>0$ or $\delta_k < 0$ in the limit. 
For example, if $\mu = 1$ then one may ensure that $c$ converges to 1 by taking any $\sigma= o(k^{-1/2})$. 
In the borderline case where $\lim_{k \to \infty} \delta_k= 0$, we still have $\lim_{k \to \infty} c= 1$ as long as $k^{-2/3}= o(\delta_k)$. 
Although Theorem \ref{thm:Wigner} is an asymptotic result, the following example illustrates that it provides a good depiction of the behavior of $c$ for finite $k$.

\begin{example}
    Fix the diagonal $\theta_{ii} = 1$ and let $\pi$ be a zero-mean Gaussian. 
    Figure \ref{fig:FixedDiag} shows an approximation to $c$ for $k \in [2,200]$, obtained via sampling from $p$. 
    For constant $\delta_k \in \{ -0.1,-0.05,0,0.05,0.1 \}$ (left panel), $c$ remains close to 1 for the two $\delta_k > 0$ settings, but quickly decreases towards 0 for the two $\delta_k < 0$ settings. For the borderline $\delta_k=0$, $c$ decreases slowly as $k$ grows. For decreasing $\delta_k \in \{ k^{-1/2},k^{-2/3},k^{-1},k^{-2} \}$ (right panel), $c$ remains close to 1 for $\delta_k = k^{-1/2} \gg k^{-2/3}$. For the two $\delta_k \ll k^{-2/3}$ settings, the values of $c$ are comparable to those under $\delta_k = 0$ (note the different y-axis scale, relative to the left panel). 
    For the borderline $\delta_k = k^{-2/3}$, $c$ slowly decreases as $k \to \infty$.

    \begin{figure}[h]\label{fig:FixedDiag}
        \caption{Monte Carlo estimate of $c=\mathbb{P}(\Theta \succ 0)$ for fixed unit diagonal and Gaussian off-diagonals for different $\delta_k= \frac{ \mu }{ \sigma \sqrt{ k } } - 2$. Left: fixed $\delta_k$. Right: $\delta_k$ varies with $k$}
	   \centering
       \begin{tabular}{cc}
	       \includegraphics[scale=0.68]{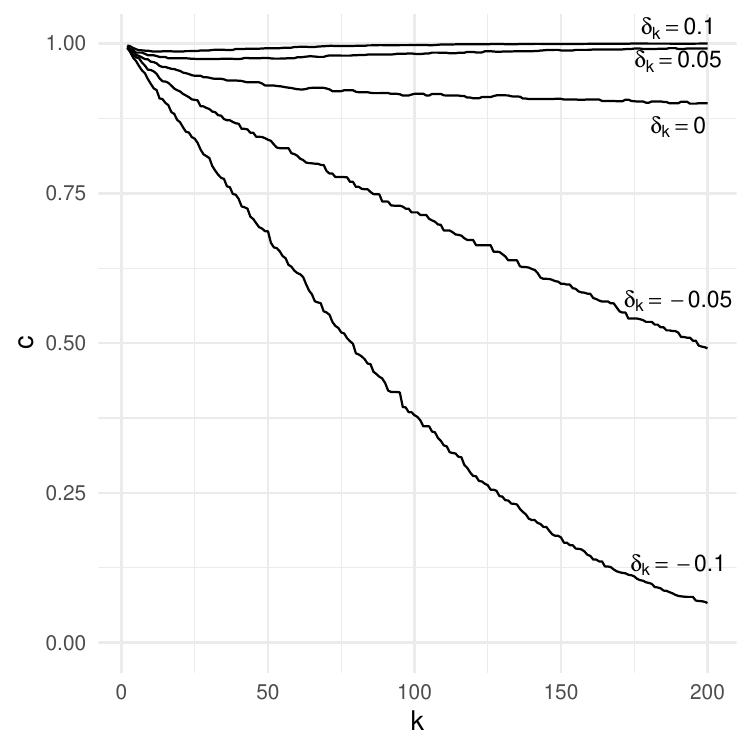} &
           \includegraphics[scale=0.68]{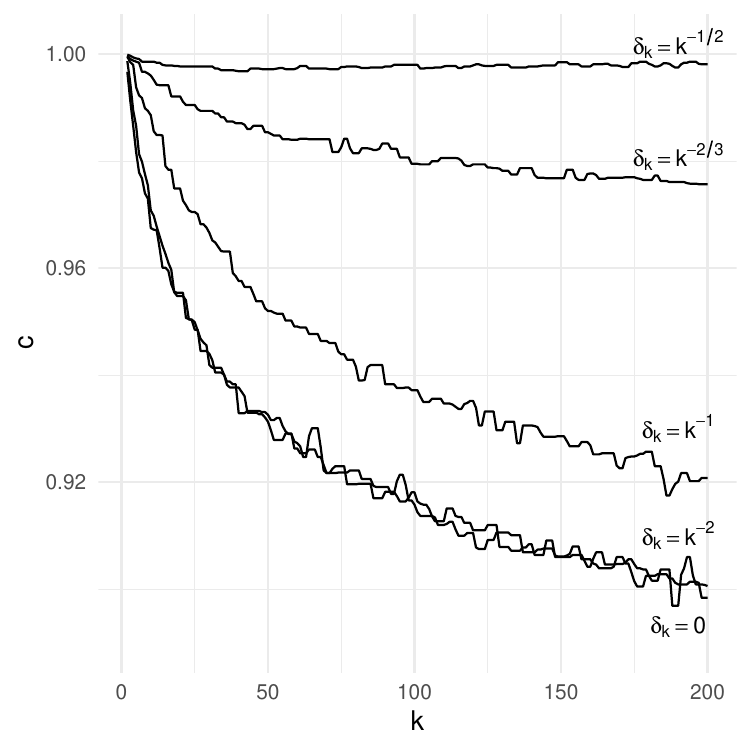}
       \end{tabular}
    \end{figure}
\end{example}

Theorem \ref{thm:Wigner2} allows the diagonal of $\Theta$ to be stochastic. The limit of $c$ then also depends on the distribution of the minimum and maximum diagonal entries, $\theta_{\min}$ and $\theta_{\max}$. 

\begin{theorem}\label{thm:Wigner2}
Let $\Theta$ be as in Theorem \ref{thm:Wigner}, but with independent stochastic diagonals $\theta_{ii}$. 
\begin{enumerate}[label=\roman*)]
    \item\label{item1} If $\lim_{k \to \infty} \mathbb{P}_p \left( \frac{ \theta_{\min} }{ \sigma \sqrt{ k } } > 2 + \epsilon \right) = 1$ for some fixed $\epsilon > 0$, then $\lim_{k \to \infty} c= 1$.
    \item\label{item2} If $\lim_{k \to \infty} \mathbb{P}_p \left( \frac{ \theta_{\max} }{ \sigma \sqrt{ k } } < 2 - \epsilon \right) = 1$ for some fixed $\epsilon > 0$, then $\lim_{k \to \infty} c = 0$.
\end{enumerate}
\end{theorem}

\begin{example}
    Let $\theta_{ij} \sim N(0,\sigma^2)$ and $\theta_{ii} \sim \mbox{Exp}(\lambda=1)$. 
    Then $\theta_{\min} \sim \mathrm{Exp}(k\lambda)$ and $\frac{\theta_{\min}}{\sigma \sqrt{k}} \sim \mathrm{Exp}(k^{3/2}\sigma\lambda)$. The condition of Theorem \ref{thm:Wigner2}(\ref{item1} is therefore satisfied if $\sigma= o(k^{-3/2})$. On the other hand, it can be shown that the condition of Theorem \ref{thm:Wigner2}(\ref{item2} is satisfied when $k^{-1/2} \log(k) = o(\sigma)$. The left panel of Figure \ref{fig:ExpGamDiag} shows $c$ for $\sigma \in \{ k^{-2},k^{-3/2},k^{-5/4},k^{-9/8},k^{-1} \}$. The setting $\sigma = k^{-2} = o(k^{-3/2})$ maintains $c$ close to 1, as does the borderline case $\sigma = k^{-3/2}$. For the remaining settings $c$ is not close to 1, and, even though the condition of Theorem \ref{thm:Wigner2}(\ref{item2} is not satisfied for any of the choices of $\sigma$, we still observe the limiting $c=0$ at $k=200$ for $\sigma= k^{-1}$.

So if $\theta_{ii}$ are exponential then $\sigma$ must be much smaller for $c$ to converge to 1 relative to the fixed diagonal case, where $\sigma= o(k^{-1/2})$ sufficed.
If instead $\theta_{ii}$ are Gamma-distributed with equal shape and rate $\alpha = \beta > 1$, then $\theta_{ii}$ has lighter left tails than the exponential distribution. It can be shown that this allows the condition on $\sigma$ to be weakened to $\sigma = o(k^{-1/2 - 1/\alpha})$. The right panel of Figure \ref{fig:ExpGamDiag} shows $c$ for $\alpha = \beta = 2$ for different settings of $\sigma$. As predicted, $c$ stays close to 1 when $\sigma = k^{-5/4}$ and $\sigma= k^{-9/8}$, as in both cases $\sigma = o(k^{-1})$, whereas $c$ is much smaller when $\sigma$ is larger than  $k^{-1}$.
    \begin{figure}[h]\label{fig:ExpGamDiag}
        \caption{Monte Carlo estimate of $c=\mathbb{P}(\Theta \succ 0)$ for $\theta_{ii} \sim \mathrm{Exp}(1)$ (left) and $\theta_{ii} \sim \mathrm{Gamma}(2,2)$ (right) and Gaussian off-diagonals with different standard deviations $\sigma$}
	   \centering
       \begin{tabular}{cc}
	       \includegraphics[scale=0.68]{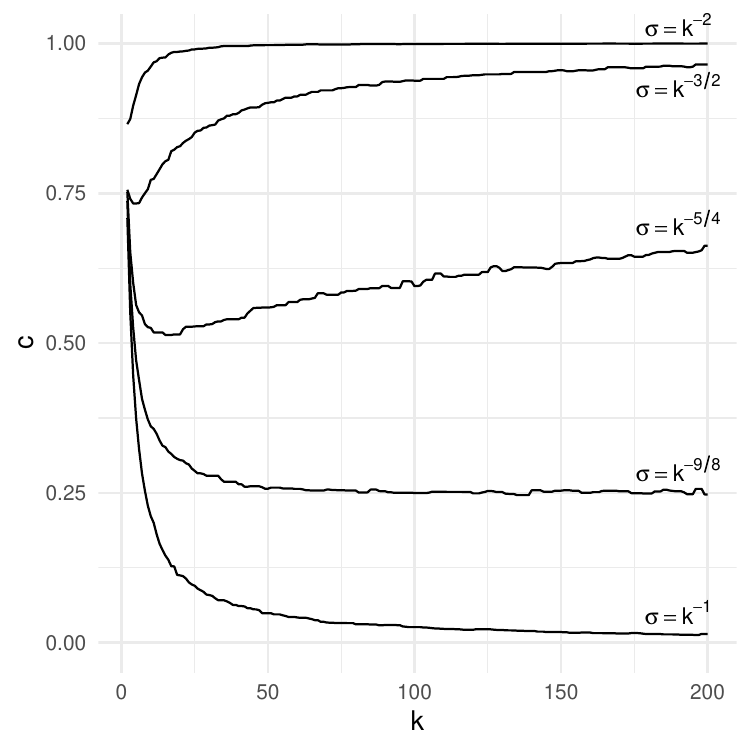} &
           \includegraphics[scale=0.68]{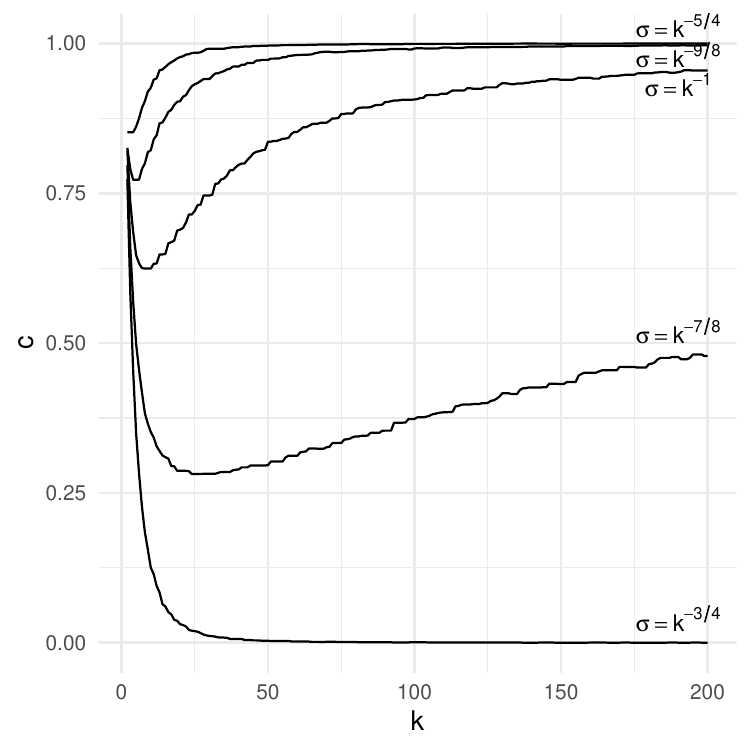}
       \end{tabular}
    \end{figure}
\end{example}

\section{Sparse matrices}
\label{sec:sparse_matrices}

An important class of priors 
are those that induce sparsity in $\Theta$. For example, if $\Theta$ is a covariance matrix then $\theta_{ij}=0$ indicates that variables $(i,j)$ are uncorrelated, and if $\Theta$ is a partial covariance matrix that their partial correlation is zero, given all other variables.
Recall that the sparsity pattern is specified by the $k \times k$ binary matrix $Z$, such that $\theta_{ij} \mid Z_{ij} = 0 \sim \delta_0$ and $\theta_{ij} \mid Z_{ij} = 1 \sim p_{ij}$ where $\delta_0$ is a Dirac mass at 0 and $p_{ij} : \mathbb{R} \to \mathbb{R}^+$ is some density function, referred to as the slab density.
A popular strategy \citep{Wang2015,Banerjee2015,Gan2018,sulem2025bayesian} is to set
\begin{align}
p^+(\Theta, z) = \frac{ p(\Theta \mid Z=z) p(z) \mathbb{I}(\Theta \succ 0)}{c}
\label{eq:jointprior_sparse}
\end{align}
where $p(z)$ is some prior on $Z$, and
$ p(\, \Theta \mid Z=z \,) $
is a separable distribution with off-diagonal marginals $p_{ij}(\theta_{ij}) \mathbb{I}(z_{ij}=1) + \delta_0(\theta_{ij})\mathbb{I}(z_{ij}=0)$.
The corresponding PD-separable distribution is 
$$
p^+(\,\Theta \mid Z=z\,) = \frac{p(\,\Theta \mid Z=z\,) \mathbb{I}(\Theta \succ 0)}{c_z},
$$
where $c_z=\mathbb P_p(\Theta \succ 0 \mid Z=z)$ is the probability of positive definiteness under the fixed sparsity pattern $z$.
As before, $c= \mathbb{P}_p(\Theta \succ 0)$ is its marginal probability.

A canonical choice for the prior on $Z$ is $p(z)= \prod_{i<j} \mbox{Bern}(z_{ij}; \eta)$, in which case the marginal $p(\Theta)$ is separable with off-diagonal margins $\pi_{ij}(\theta_{ij}) = \eta \, p_{ij}(\theta_{ij}) + (1 - \eta) \delta_0(\theta_{ij})$. The corresponding PD-separable distribution is $p^+(\Theta) = \sum_z p^+(\Theta,z)$ and we denote its margins by $\pi^+_{ij}(\theta_{ij})$. If there is also a common slab density, $p_{ij} = p_s$ for all $i<j$, then $p(\Theta)$ has common off-diagonal margins $\pi_{ij} = \pi$ for all $i<j$, as does $p^+(\Theta)$, $\pi^+_{ij} = \pi^+$ for all $i<j$.



We remark that \eqref{eq:jointprior_sparse} implies the marginal prior on the sparsity structure $p^+(z) \propto c_z p(z)$. As we discuss below this is computationally convenient, however a concern is that $p^+(z)$ may differ significantly from $p(z)$ when some of the $c_z$'s are close to 0. For example, for $p(z)= \prod_{i<j} \mbox{Bern}(z_{ij}; \eta)$, setting large $\eta \in [0,1]$ results in a non-sparse $p(z)$, but $p^+(z)$ could still be sparse if $c_z$ is small for non-sparse $z$'s.
To address this issue, an alternative to the joint prior in \eqref{eq:jointprior_sparse} is to start by defining the marginal $\tilde{p}^+(z)=p(z)$, and subsequently set $\tilde{p}^+(\Theta \mid Z=z) = p^+(\Theta \mid Z=z)$ as before, defining the alternative joint prior
\begin{align}
\tilde{p}^+(\Theta, z)= \tilde{p}^+(\Theta \mid Z=z) \tilde{p}^+(z)
= p^+(\Theta \mid Z=z) p(z).
\label{eq:jointprior_sparse_alternative}
\end{align}

While \eqref{eq:jointprior_sparse_alternative} may feel more natural, in that one directly sets a prior on the sparsity structure $\tilde p^+(z)=p(z)$, this strategy faces a doubly-intractable problem where both prior and posterior normalisation constants are unavailable.
Specifically, given some data $Y$ with associated likelihood $p(Y \mid \Theta)$, the posterior distributions on $Z$ under 
\eqref{eq:jointprior_sparse} and \eqref{eq:jointprior_sparse_alternative} are, respectively,
\begin{align}
&p^+(z \mid Y) \propto p(Y \mid Z=z) p^+(z)= \frac{c_z p(z)}{c_z} \int p(Y \mid \Theta, Z=z) \mathbb I(\Theta \succ 0) d P(\Theta \mid Z=z)
\nonumber \\
&\tilde{p}^+(z \mid Y) \propto p(Y \mid Z=z) p(z)= \frac{p(z)}{c_z} \int p(Y \mid \Theta, Z=z) \mathbb I(\Theta \succ 0) d P(\Theta \mid Z=z)= \frac{p^+(z \mid Y)}{c_z}.
\label{eq:postprob_z}
\end{align}

Whereas $\tilde{p}^+(z \mid Y)$ requires evaluating $c_z$, which is computationally-prohibitive when one considers many $z$'s, $p^+(z \mid Y)$ associated to \eqref{eq:jointprior_sparse} does not.
Therefore one sets the prior in \eqref{eq:jointprior_sparse} out of computational necessity, but this has an effect on posterior inference. Specifically, the ratio of posterior masses on two sparsity structures $(z,z')$ satisfies
$$
\frac{p^+(z \mid Y)}{p^+(z' \mid Y)}= \frac{c_z}{c_{z'}} \frac{\tilde{p}^+(z \mid Y)}{\tilde{p}^+(z' \mid Y)}.
$$


Therefore, it is of interest to study how $p^+(z) \propto c_z p(z)$ differs from $p(z)$.
To do so, in Section \ref{ssec:control_cz} we show that $p^+(z)$ favors sparser models relative to $p(z)$, and that this effect becomes more marked as the prior variance on off-diagonal $\theta_{ij}$'s grows.
We also obtain bounds for $c_z$ and related TV distance bounds between $p^+(z)$ and $p(z)$.
Section \ref{sec:control_c_sparse} describes the behavior of $c$ by extending results from the dense case in Section \ref{sec:contolc}.
This is relevant because, by the same logic as Proposition \ref{prop:dist}, $c$ 
determines the distance between 
$p^+(\Theta,z)$ and $p(\Theta,z)$. Hence $c$ also bounds the distance between marginals $p^+(\Theta)$ and $p(\Theta)$, and between $p^+(z)$ and $p(z)$.
Additionally, $c = \sum_z c_z p(z)$ and so having $c \approx 1$ ensures that $c_z \approx 1$ for any $z$ with large $p(z)$. 
If $p(z)= \prod_{i < j} \mbox{Bern}(z_{ij}; \eta)$
(as will be assumed in Section \ref{sec:control_c_sparse}) then $p^+(\Theta)$ is PD-separable
and by Proposition \ref{prop:margindist_upperbound} we have that $c$ bounds the TV and KL distances between the marginals $\pi^+(\theta_{ij})$ and $\pi(\theta_{ij})$.
Note that $\mbox{TV}(\pi^+, \pi)$ by definition bounds $|\mathbb P_{\pi^+}(\theta_{ij} \neq 0) - \mathbb P_{\pi}(\theta_{ij} \neq 0)|$, that is the difference between critical measures of prior sparsity.

\subsection{Controlling $c_z$}
\label{ssec:control_cz}

As discussed, introducing $c_z= \mathbb P_p(\Theta \succ 0 \mid Z=z)$ into the model prior as $p^+(z) \propto c_z p(z)$ can affect posterior inference, particularly when some $c_z \approx 0$.
We now study the behavior of $c_z$. The focus is on how the sparsity in $Z=z$ and the prior off-diagonal variance $\sigma^2$ influence $c_z$, and on bounding $c_z$.
One of our main results is Proposition \ref{prop:mean_eigenval_z}. It shows that, as $z$ becomes denser, then $\lambda_{\min}(\Theta)$ decreases in expectation (when the $\theta_{ij}$'s arise from Gaussians or scale mixtures of Gaussians). Intuitively, this suggests that $c_z$ decreases as $z$ gets denser, that is $p^+(z \mid Y)$ places higher posterior probability on sparse models than $\tilde{p}^+(z \mid Y)$.
Corollary \ref{cor:cbounds_sparse} complements Proposition \ref{prop:mean_eigenval_z} by showing that lower and upper bounds for $c_z$ indeed decrease as $z$ gets denser.
Corollary \ref{cor:sigma_sparse} shows that $c_z$ strictly decreases from 1 to 0 as $\sigma$ ranges from 0 to $\infty$, extending Proposition \ref{prop:sigma} to the sparse setting. 
Our other main result is Theorem \ref{thm:sparse_wigner_gaussian}, 
which uses random matrix theory to obtain much sharper lower bounds on $c_z$ than Corollary \ref{cor:cbounds_sparse}.
Corollary \ref{cor:tvd_sparse} shows that such lower bounds allow controlling the TV distance between the priors $p^+(z)$ and $\tilde{p}(z)$.

Proposition \ref{prop:mean_eigenval_z} shows that, when the slab densities $p_{ij}$ are Gaussian or mixtures of Gaussians, the minimum eigenvalue $\lambda_\min(\Theta)$ increases in expectation as $z$ becomes sparser (has more zero entries).
The result for mixtures allows having a common $\sigma_{ij}=\sigma$ with probability 1, which for example can be used to induce a marginal multivariate t distribution on $\Theta$, as well as different $\sigma_{ij}$'s, for example inducing independent univariate t marginal distributions on each $\theta_{ij}$.

\begin{proposition}\label{prop:mean_eigenval_z}
    Suppose that either $p_{ij}(\theta_{ij})= N(\theta_{ij}; 0, \sigma_{ij}^2)$ independently across $i < j$ or, more generally, that for some $\Pi(\{\sigma_{ij}^2\}_{i<j})$,
    $$p(\{\theta_{ij}\}_{i<j})= \int \prod_{i < j} N(\theta_{ij}; 0, \sigma_{ij}^2) d\Pi(\{\sigma_{ij}^2\}_{i<j})$$
    For two fixed sparsity patterns $z$ and $z'$ with $z_{ij} \leq z_{ij}'$, $$\mathbb E_p[\, \lambda_\min(\Theta) \mid Z = z \,] \geq \mathbb E_p[\, \lambda_\min(\Theta) \mid Z = z' \,]$$
    For random sparsity pattern with $Z_{ij} \sim \mathrm{Bern}(\eta_{ij})$, $\mathbb E_p[ \lambda_\min(\Theta) ]$ is decreasing in each $\eta_{ij}$.
\end{proposition}

Lemma \ref{lem:eigenvalue_concentration} shows that, given $Z=z$, $\lambda_{\min}(\Theta)$ concentrates on its expectation (in the Gaussian case).
For simplicity, we focus on the case where $\Theta$ has a fixed diagonal.

\begin{lemma}
Suppose that $\theta_{ii}=a$ for some deterministic $a>0$ and that $p_{ij}(\theta_{ij})= N(\theta_{ij}; 0, \sigma_{ij}^2)$ independently across $i < j$. Then
$\mathbb P \left( |\lambda_{\min}(\Theta) - \mathbb E[\lambda_{\min}(\Theta) \mid Z=z] | \geq t \sigma \mid Z=z \right) \leq e^{-t^2/2}$.

 
\label{lem:eigenvalue_concentration}
\end{lemma}

\begin{corollary} \label{cor:cbounds_sparse}
    Assume the settings in Proposition \ref{prop:cbounds}, but with sparsity pattern $Z=z$. Then
    \begin{align*}
        \mathbb{E}\left[ \left( 2 \Phi \left( \theta_{\min}/(k-1) \sigma \right) - 1 \right)^{\sum_{i<j}z_{ij}} \right] \leq c_z \leq \mathbb{E}\left[\left( 2 \Phi \left( \theta_{\max}/\sigma \right) - 1 \right)^{\sum_{i<j}z_{ij}}\right]
    \end{align*}
    where the expectations are with respect to $\theta_{\min}= \min_{i=1,\ldots,k} \theta_{ii}$ and $\theta_{\max}= \max_{i=1,\ldots,k} \theta_{ii}$.

\end{corollary}

Corollary \ref{cor:cbounds_sparse} follows from Proposition \ref{prop:cbounds},
with the difference that the exponent is now $\sum_{i<j}z_{ij} \leq k(k-1)/2$. Both the lower and upper bounds for $c_z$ decrease as $\sum_{i<j} z_{ij}$ grows, that is as $z$ gets denser.

\begin{corollary} \label{cor:sigma_sparse}
Assume the settings in Proposition \ref{prop:sigma}, but with non-empty sparsity pattern $Z=z$.
Then, for any fixed diagonal distributions $\pi_{ii}$, $c_z$ is continuous and strictly decreasing in $\sigma$. Further, $\lim_{\sigma \to 0} c_z= 1$ and $\lim_{\sigma \to \infty} c_z= 0$.
\end{corollary}

We next obtain sharper lower bounds on $c_z$.
For context, \cite{van2017structured} (Example 4.9) showed an asymptotic result for the case where $\theta_{ii} \sim N(\mu, \sigma^2)$ and $\theta_{ij} \mid Z_{ij} = 1 \sim N(0, \sigma^2)$ independently across $(i,j)$.
Specifically, if $\mu \geq a \sigma (\sqrt{d_z} + \sqrt{\log k})$ for some constant $a>0$ then $\lim_{k \to \infty} \mathbb E_p(\lambda_{\min}(\Theta) \mid Z=z)>0$, where recall that $d_z= \max_i\sum_{j \neq i} z_{ij}$ is the maximum degree of $z$.
Although in a Bayesian framework it seems unnatural to set a Gaussian prior assigning positive probability to diagonal $\theta_{ii}<0$, this result suggests that in sparse settings $\lambda_{\min}(\Theta)$ scales much better with dimension than in the dense setting, where $\mu \geq 2 \sigma \sqrt{k}$ is required (Theorem \ref{thm:Wigner}) to guarantee that $\lambda_{\min}(\Theta)>0$ (and hence $c_z=1$) in the limit as $k \to \infty$.
Theorem \ref{thm:sparse_wigner_gaussian} makes this intuition rigorous, allowing for $\theta_{ii}$ to be either fixed or exponentially-distributed. Besides limiting results, it gives exact lower-bounds for $c_z$ for each fixed $(k,d_z)$. 

Part (i) refers to the case where $\Theta$ has fixed diagonal $\mu$ and states that $c_z$ is large when $\mu^2/(2\sigma^2 d_z) - \log k$ is small. Relative to Corollary \ref{cor:cbounds_sparse}, the dependence on $k$ is logarithmic and the dependence on $z$ via its largest degree $d_z$ rather than the number of non-zero entries in $z$.
Part (ii) extends Part (i) beyond the Gaussian setting.
The idea is that one truncates $|\theta_{ij}| \leq a$ for some $a \leq \mu$, a mild condition since $|\theta_{ij}| \leq \mu$ is necessary for $\Theta \succ 0$. In particular, for correlation and partial correlation matrices with unit diagonal one may take $a=1$.
Extensions to non-truncated $\theta_{ij}$'s are possible using extended Bernstein inequalities (\cite{tropp:2012}, Section 6) but we do not consider them here for brevity. 
Part (iii) considers the case where diagonal $\theta_{ii}$ are exponentially-distributed with mean $\mu=1/\lambda$.
Denote by $r= \sigma \lambda= \sigma/\mu$, 
neglecting log terms, Part (iii) states if $r$ is small then $c_z$ is large. For example, if $r = o(1/[k \sqrt{d_z}])$ then $c_z$ converges to 1, a significantly stronger requirement than the $r= o(1/\sqrt{d_z})$ required by Part (i). That is, when the diagonal is fixed one can control $c_z$ under much milder conditions on $\sigma$.

\begin{theorem}
Assume that $\theta_{ii}$ (when non-deterministic) and $\theta_{ij}$ are mutually independent for $i \geq j$.
Let $d_z$ be the maximum degree associated to the binary $k \times k$ matrix $z$.

\begin{enumerate}[leftmargin=*]

\item[(i)]
If $\theta_{ii}=\mu$ almost surely for $i=1,\ldots,k$ and $\theta_{ij} \mid Z_{ij}=1 \sim N(0,\sigma^2)$, then $c_z \geq 1 - 2 k e^{-\mu^2/(2 \sigma^2 d_z)}$.
Hence, if $\mu \geq \sigma \sqrt{2 d_z \log(2k/\alpha)}$ then $c_z \geq 1 - \alpha$.

\item[(ii)] If $\theta_{ii}= \mu$ and $|\theta_{ij}| \leq a$ for some $a>0$, almost surely with $\mathbb E(\theta_{ij} \mid Z_{ij}=1)=0$ and $Var(\theta_{ij} \mid Z_{ij}=1)=\sigma^2$, then
\begin{align}
c_z \geq 1 - 2 k \exp \left\{ - \frac{\mu^2}{2 (\sigma^2 d_z + a \mu/3)} \right\}.
\nonumber
\end{align}
Further, if $\mu \geq 1$ and $\mu \geq 2 (\sigma^2 d_z + a/3) \log(2k/\alpha)$ then $c_z \geq 1-\alpha$.

\item[(iii)] Let $\theta_{ii} \sim \mbox{Exp}(1/\mu)$ and $\theta_{ij} \mid Z_{ij} = 1 \sim N(0,\sigma^2)$. Then
$c_z \geq 1 - e^{- k t / \mu} + 2 k e^{-\frac{t^2}{2 \sigma^2 d_z}}$
for any $t>0$.
In particular, assume that $ \sigma \sqrt{d_z} / \mu \leq 2\sqrt{2}$, then
\begin{align}
c_z \geq 1 -   \frac{\sqrt{2} k \sigma \sqrt{d_z}}{\mu} \left[ \sqrt{\ln\left(\frac{2\sqrt{2} \mu}{\sigma \sqrt{d_z}}\right)} + 1 \right].
\nonumber
\end{align}

Further, if 
$\mu \geq \frac{2 \sigma k \sqrt{d_z}}{\alpha} \sqrt{- W_{-1}\left( \frac{- \alpha^2}{32 e^2 k^2} \right)}$
then $c_z \geq 1 - \alpha$, where $W_{-1}()$ is the branch of Lambert's $W_q$ function associated to $q=-1$.

\end{enumerate}

\label{thm:sparse_wigner_gaussian}
\end{theorem}


All these lower bounds are increasing in the largest degree $d_z$. 
Corollary \ref{cor:tvd_sparse} states that these bounds directly translate into the TV distance between $p^+(\Theta)$ and $p(\Theta)$, which in turn bound that between $\pi^+(\theta_{ij})$ and $\pi(\theta_{ij})$ (for separable $p(\Theta)$).
This result is relevant in sparse settings where one constrains the maximum degree $d_z \leq \bar d$ for some $\bar d = o(k)$.

\begin{corollary}   
\label{cor:tvd_sparse}
Suppose that $c_z \geq 1 - b(\mu,\sigma^2,d_z)$ for some bound $b()$ that is decreasing in $d_z$, and that $p(z) = 0$ for any $z$ with largest degree $d_z > \bar d$. Then $c \geq 1 - b(\mu, \sigma^2, \bar d)$ and the TV and KL distance between 
$p^+(\Theta)$ and $p(\Theta)$
satisfy
$\mbox{TV}(p^+, p) \leq b(\mu,\sigma^2,\bar d)$
and $KL(\,p^+ \;||\; p\,) \leq - \log (1 - b(\mu,\sigma^2, \bar d))$.
Further, suppose that $p(z)= \prod_{i < j} \mbox{Bern}(z_{ij}; \eta)$  for some $\eta \in [0,1]$ 
and there is a common slab density $p_{ij}(\theta_{ij}) = p_S(\theta_{ij})$ for all $i<j$. 
Then the TV and KL distances between off-diagonal $\pi(\theta_{ij})$ and $\pi^+(\theta_{ij})$ satisfy
\begin{align*}
&KL(\,\pi^+ \;||\; \pi\,) \leq - \frac{2\log(1 - b(\mu,\sigma^2, \bar d))}{k(k-1)}
\\
&\mbox{TV}( \pi^+, \pi) 
\leq \min\left\{ \sqrt{\frac{-\log(1 - b(\mu,\sigma^2, \bar d))}{k(k-1)}}, \sqrt{ 1 - e^{\frac{2 \log(1 - b(\mu,\sigma^2, \bar{d}))}{k(k-1)}} } \right\}
.
\end{align*}
\end{corollary}
Note that 
Corollary \ref{cor:tvd_sparse} bounds the difference in marginal inclusion probabilities
$
\abs{\mathbb{P}_p(Z_{ij} = 1) - \mathbb{P}_{p^+}(Z_{ij} = 1)} =\abs{\mathbb{P}_p(\theta_{ij} \neq 0) - \mathbb{P}_{p^+}(\theta_{ij} \neq 0)} \leq \mbox{TV}( \pi^+, \pi).
$
This is important because $\mathbb{P}_{p^+}(Z_{ij} = 1)$ is a natural measure of prior sparsity.
By combining the bounds in Corollary \ref{cor:tvd_sparse} with those of Theorem \ref{thm:sparse_wigner_gaussian}, one can derive conditions for which $\mbox{TV}( \pi^+, \pi ) \to 0$ as $k \to \infty$.
For convenience, we give these in Corollary \ref{cor:tvd_sparse_suffcond} and discuss the implications below.


\begin{corollary} \label{cor:tvd_sparse_suffcond}
Suppose that $p(z)= \prod_{i < j} \mbox{Bern}(z_{ij}; \eta)$  for some $\eta \in [0,1]$ and there is a common slab density $p_{ij}(\theta_{ij}) = p_S(\theta_{ij})$ for all $i<j$. 
Denote by $\mbox{TV}^\infty(p^+, p)= \lim_{k \to \infty} \mbox{TV}(p^+,p)$
and by $\mbox{TV}^\infty(\pi^+, \pi)= \lim_{k \to \infty} \mbox{TV}(\pi^+,\pi)$.
\begin{enumerate}[leftmargin=*]
\item[(i)] Assume the conditions of Theorem \ref{thm:sparse_wigner_gaussian}(i).
If $\lim_{k \to \infty} \mu^2/(2 \sigma^2 \bar d) - \log(k)= \infty$,
then $\mbox{TV}^\infty(p^+,p)=0$.
Further, if 
$\lim_{k \to \infty} \mu^2 / (2 \sigma^2 \bar d) - \log(k) > \log 2$
then $\mbox{TV}^\infty(\pi^+,\pi)=0$.

\item[(ii)] Assume the conditions of Theorem \ref{thm:sparse_wigner_gaussian}(ii).
If $\frac{a\mu/3}{\sigma^2 \bar d + a\mu/3}=o(1)$, then the results from Part (i) hold replacing $\mu^2$ by $(1-\epsilon) \mu^2$, where $\epsilon >0$ can be taken arbitrarily close to 0.

\item[(iii)] Assume the conditions of Theorem \ref{thm:sparse_wigner_gaussian}(iii).
If $\mu/(\sigma \sqrt{\bar d}) \geq k s_k$ for any $s_k$ satisfying $\sqrt{\log k} = o(s_k)$, then $\mbox{TV}^\infty(p^+,p)=0$.
Further, if $\mu/(\sigma \sqrt{\bar d}) = \sqrt{2} k s_k$ where $s_k \geq \sqrt{\log k} + 2$, then $\mbox{TV}^\infty(\pi^+,\pi)=0$.
\end{enumerate}

\end{corollary}

An implication is that if the diagonal entries are fixed, as in Parts (i)-(ii),  then one can guarantee that $\mbox{TV}^\infty(p,p^+)=0$ and $\mbox{TV}^\infty(\pi^+,\pi)=0$ under much milder conditions than when they are exponentially-distributed (Part (iii)).
For example, Parts (i)-(ii) guarantee that $\mbox{TV}(p^+,p)$ vanishes if $\mu/(\sigma \sqrt{\bar d}) = \sqrt{2 \log k} + s_k$ for some $s_k$ such that $\lim_{k \to \infty} s_k = \infty$,
whereas Part (iii) requires that  $\mu/(\sigma \sqrt{\bar d})$ grows at a strictly larger rate than $k \sqrt{\log k}$, which is a much worse dependence on the dimension $k$.
Another implication is that, as discussed before Proposition \ref{prop:margindist_upperbound}, the conditions for $\mbox{TV}^\infty(\pi^+,\pi)=0$ are slightly milder than those for $\mbox{TV}^\infty(p^+,p)=0$.
In Part (i) for example, $\mbox{TV}^\infty(\pi^+,\pi)=0$ if $\mu/\sigma \geq \sqrt{2 \bar d [\log(k) + \epsilon]}$ with $\epsilon > \log 2$, relative to the $\mu/\sigma= \sqrt{2 \bar d \log k} + s_k$ with $s_k \to \infty$ required by $\mbox{TV}^\infty(p^+,p)$.


\subsection{Controlling $c$ in the sparse setting}
\label{sec:control_c_sparse}

Corollaries \ref{cor:SparseWigner}-\ref{cor:SparseWigner_randomdiag} describe the limit of $c= \mathbb P_p(\Theta \succ 0)$, by extending Theorems \ref{thm:Wigner}-\ref{thm:Wigner2} to the sparse setting. This quantity remains important in the sparse setting as it is the normalising constant of the joint distribution in (\ref{eq:jointprior_sparse}) and $c = \sum_z c_z p(z)$ and so having $c \approx 1$ ensures that $c_z \approx 1$ for all $z$ with large $p(z)$.

\begin{corollary}\label{cor:SparseWigner}
Let $\theta_{ii}=\mu$ for some deterministic $\mu>0$ and $\theta_{ij}= \sigma Z_{ij} \tilde{\theta}_{ij}$ for $1 \leq i < j \leq k$, where $\tilde{\theta}_{ij}$ are i.i.d. variables with zero mean, unit variance, and finite fourth moment,
and $Z_{ij} \sim \mbox{Bern}(\eta_k)$ independently across $i<j$ with $\eta_k$ satisfying $k^{-1/3}= o(\eta_k)$.

    \begin{enumerate}[label=\roman*)]
        \item If $\liminf_{k \rightarrow \infty} \frac{ \mu }{ \sigma \sqrt{ \eta_k k } } > 2$, then $\lim_{k \to \infty} c = 1$.
        \item If $\limsup_{k \rightarrow \infty} \frac{ \mu }{ \sigma \sqrt{ \eta_k k } } < 2$, then $\lim_{k \to \infty} c = 0$.
    \end{enumerate}
    Further, if one sets any $\sigma = \frac{\mu}{(2+\delta_k) \sqrt{\eta_kk}}$ where $\delta_k$ satisfies $k^{-2/3}= o(\delta_k)$, then $\lim_{k \to \infty} c= 1$.
\end{corollary}

\begin{corollary} \label{cor:SparseWigner_randomdiag}
    Let $\Theta$ be as in Corollary \ref{cor:SparseWigner} but with independent stochastic diagonal entries $\theta_{ii}$. Let $\theta_{\min} = \min_{i=1,\dots,k}\theta_{ii}$ and $\theta_{\max} = \max_{i=1,\dots,k} \theta_{ii}$.
\begin{enumerate}[label=\roman*)]
    \item If $\mathbb{P}\left( \frac{ \theta_{\min} }{ \sigma \sqrt{ \eta_k k } } > 2 + \epsilon \right) \to 1$ for some $\epsilon > 0$, then $\lim_{k \to \infty} c= 1$.
    \item If $\mathbb{P}\left( \frac{ \theta_{\max} }{ \sigma \sqrt{ \eta_k k } } < 2 - \epsilon \right) \to 1$ for some $\epsilon > 0$, then $\lim_{k \to \infty} c = 0$.
\end{enumerate}
\end{corollary}

It is important to note the condition $k^{-1/3} = o(\eta_k)$. For sparser regimes where $\eta_k= O(k^{-1/3})$, the standard Wigner matrix limit of the extreme eigenvalues no longer holds. 
We demonstrate this with an example.
\begin{example}
    Let $\theta_{ii}=1$ and $\theta_{ij}= \sigma Z_{ij} \tilde{\theta}_{ij}$ where 
    $\tilde{\theta}_{ij} \sim N(0,1)$. We consider $\sigma = \frac{1}{(2+\delta_k)\sqrt{k\eta_k}}$ with $\delta_k \in \{ -0.1, -0.05, 0, 0.05, 0.1 \}$ and different settings of $\eta_k$. 
    The left panel in Figure \ref{fig:Sparse} demonstrates the result of Corollary \ref{cor:SparseWigner}. Taking $\eta_k = 0.5 k^{-1/4}$, $c$ converges to 1 for $\delta_k > 0$, and to $0$ when $\delta_k < 0$. 
    In the case not covered by Corollary \ref{cor:SparseWigner} where $\eta_k = 0.5 k^{-1/2} \ll k^{-1/3}$ (right panel), $c$ is close to 0 even when $\delta_k > 0$.

    \begin{figure}[h]\label{fig:Sparse}
        \caption{Monte Carlo estimate of $c= \mathbb{P}(\Theta \succ 0)$ in the sparse case for fixed $\theta_{ii}=1$ and sparse Gaussian off-diagonals for different sparsity levels $\eta_k= \mathbb{P}_p(\theta_{ij}=0)$}
	   \centering
       \begin{tabular}{cc}
       $\eta_k = 0.5 k^{-1/4}$ & $\eta_k = 0.5 k^{-1/2}$ \\
       \includegraphics[scale=0.68]{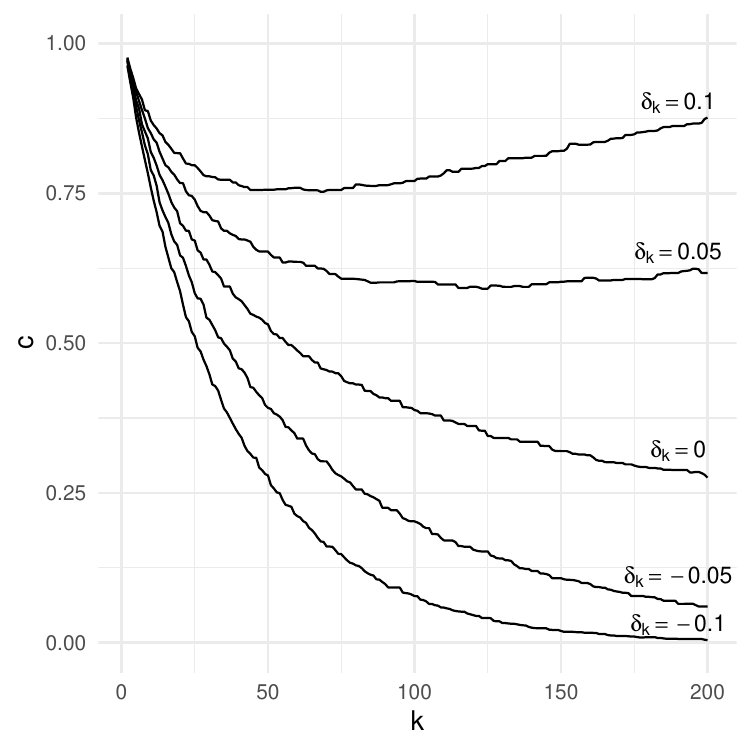} &
       \includegraphics[scale=0.68]{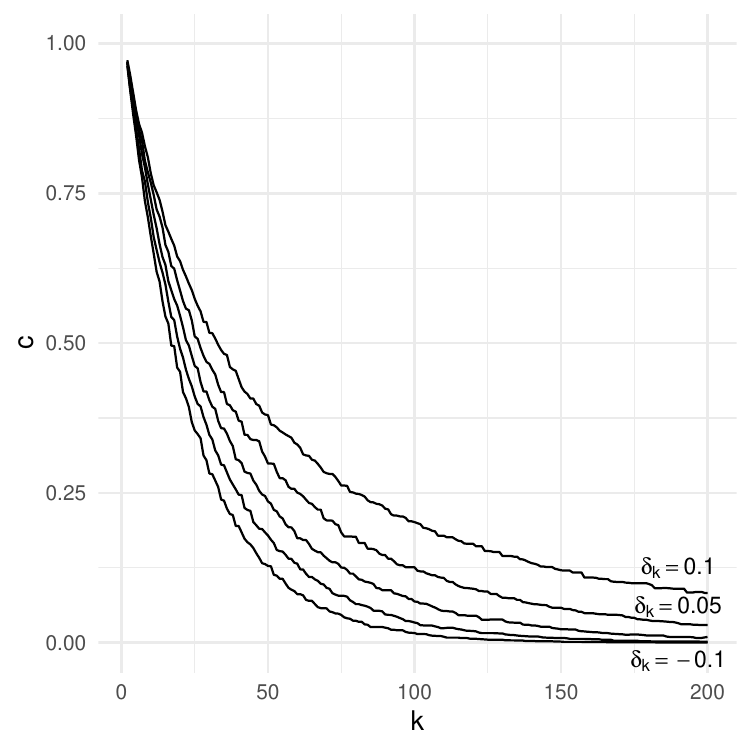}
       \end{tabular}
    \end{figure}
\end{example}

\section{Discussion}

We discussed strategies to set the prior parameters of a PD-separable prior $p^+$ such that $c= \mathbb P_p(\Theta \succ 0)$, and $c_z= \mathbb P_p(\Theta \succ 0 \mid Z=z)$ in sparse settings, are provably close to 1.
Specifically, the off-diagonal variance must be small enough relative to the minimum diagonal.
Such prior parameter values then ensure that the prior is interpretable, in the sense that $p^+$ is close to the untruncated $p$.
We also showed how one can allow for $c$ to vanish as $k$ grows and still guarantee that the marginals $\pi_{ij}^+$ converge to $\pi_{ij}$. Finally, we showed that $c_z$ plays a role in posterior inference, specifically that $c_z$ decreases as one adds non-zero entries to $z$. We also showed that $c_z$ is decreasing in $\sigma$, and developed bounds for $c_z$ that involve $\sigma$, which can be useful in setting default values for this important prior parameter.

An interesting design choice when setting $p$ is in the diagonal density. We saw that the distribution of the minimum diagonal entry is important in lower bounding the value of $c$ and $c_z$. This would suggest that a diagonal density with support on $(\epsilon,\infty)$ with $\epsilon > 0$ might be a good choice, since this ensures that the minimum diagonal is at least equal to $\epsilon$. This could be achieved by either shifting or truncating a standard density on $(0,\infty)$. If, however, it is desired that the diagonals truly have support on $(0,\infty)$, then densities with lighter tails around 0 should be preferred. This suggests replacing the Exponential diagonals, currently favored in Gaussian graphical models \citep{Wang2012,Gan2018,Jewson2022,sulem2025bayesian}, with a lighter tailed Gamma or Inverse-Gamma.

One has even a tighter control on $c$ and $c_z$ when the diagonal of $\Theta$ is fixed. This allows one to set less stringent conditions on $\sigma$ and still guarantee that $c \approx 1$ and $c_z \approx 1$.
This suggests that it may be interesting to parameterize $\Theta= D^{1/2} \widetilde{\Theta} D^{1/2}$, where $D$ is a diagonal matrix with positive diagonal, and set a PD-separable prior on $\widetilde{\Theta}$.
For example, in the context of Gaussian graphical models, $\widetilde{\Theta}$ contains partial correlations, which may be more interpretable than the partial covariances in $\Theta$. \cite{carter2024partial} showed that this parameterisation is essentially the only one enjoying a scale-invariant property, adding to its appeal.

\section{Acknowledgments}

JSC was supported by the EUTOPIA Science and Innovation Fellowship Programme and funded by the European Union Horizon 2020 programme under the Marie Skłodowska-Curie grant agreement No 945380. DR was funded by grant 2025 ICREA 00045 (Dept. of Research and Universities of the Government of Catalonia, Academy of Excellence Programme),
grant PID2022-138268NB-I00 by MCIN/AEI/10.13039/501100011033 / FEDER (EU),
and grant Consolidaci\'on investigadora CNS2022-135963 by the AEI.

\newpage
\bibliographystyle{plainnat}
\bibliography{Wigner}

\newpage

\appendix

\section{Section \ref{sec:charact} proofs}

\subsection{Proof of Proposition \ref{prop:dist}}

Let $\calS$ be the set of symmetric matrices and $\calS^+$ the set of PD matrices. The TV distance is given by
\begin{align*}
     \mbox{TV}(p,p^+) = \sup_{A \subseteq \calS} \abs{p(A) - p^+(A)}.
\end{align*}
The supremum is achieved by taking any $A$ such that $\{ \Theta: p(\Theta) < p^+(\Theta) \} \subseteq A \subseteq \{ \Theta: p(\Theta) \leq p^+(\Theta) \}$, provided that $A$ is measurable. If $\Theta \in \mathcal{S}^+$ then $p^+(\Theta) = \frac{1}{c}p(\Theta) \geq p(\Theta)$, since $c \in [0,1]$. If $\Theta \notin \mathcal{S}^+$ then $p^+(\Theta) = 0 \leq p(\Theta)$. Hence we can take $A = \calS^+$ and we have
\begin{align*}
    \mbox{TV}(p,p^+) &= \abs{p(\calS^+)-p^+(\calS^+)} 
        = \abs{c-1}
        = 1-c.
\end{align*}
Letting $q^+$ be any distribution on $\calS^+$, we have that
\begin{align*}
    \mbox{TV}(p,q^+) \geq \abs{p(\calS^+)-q^+(\calS^+)} 
        = 1-c.
\end{align*}
The Kullback-Leibler divergence between $p^+$ and $p$ is $KL(\,p^+ \;||\; p\,)=$
\begin{align*}
    \int_{\calS} p^+(\Theta) \log\left( \frac{p^+(\Theta)}{p(\Theta)} \right) \,d\Theta 
    = \int_{\calS^+} p^{+}(\Theta) \log\left( \frac{p(\Theta)}{c \, p(\Theta)} \right) \,d\Theta 
    = \log \left( \frac{1}{c} \right) \int_{\calS^+} p^{+}(\Theta) \,d\Theta 
    = -\log( c ).
\end{align*}
For any $q^+$ we have
\begin{align*}
    KL(\,q^+ \;||\; p\,)=
    KL(\,q^+ \;||\; p^+\,) + \int_\calS q^+(\Theta) \log\left( \frac{p^+(\Theta)}{p(\Theta)} \right) \,d\Theta = KL(\,q^+ \;||\; p^+\,) - \log(c).
\end{align*}
Since $KL(\,q^+ \;||\; p^+\,) \geq 0$, we have $KL(\,q^+ \;||\; p\,) \geq KL(\,p^+ \;||\; p\,)$.

\subsection{Proof of Proposition \ref{prop:marg}}

Let $\Theta_{-ij}$ denote the entries of $\Theta$ without $\theta_{ij}$. Then we have
\begin{align*}
    \pi^+_{ij}(x) &= \int_{\Theta \in \calS^+:\theta_{ij}=x} p^+\left(\Theta\right) \, d\Theta_{-ij} 
    = \frac{1}{c} \int_{\Theta \in \calS^+:\theta_{ij}=x} p\left(\Theta\right) \, d\Theta_{-ij} \\
    &= \frac{1}{c} \, \pi_{ij}(x) \int_{\Theta \in \calS^+:\theta_{ij}=x} p\left(\Theta_{-ij}\right) \, d\Theta_{-ij} 
    = \frac{1}{c}\,\pi_{ij}(x) \int_{\Theta \in \calS^+} p\left(\,\Theta \mid \theta_{ij} = x \,\right) \, d\Theta_{-ij} 
    = \frac{c_{ij}(x)}{c}\,\pi_{ij}(x).
\end{align*}

Further,
\begin{align*}
    \mathbb{E}_{\pi_{ij}}[c_{ij}(X)] &= \int c_{ij}(x) \pi_{ij}(x) \, dx 
    = \int \mathbb{P}(\,\Theta \succ 0 \mid \theta_{ij} = x \,) \pi_{ij}(x) \, dx 
    = \mathbb{P}(\,\Theta \succ 0 \,)
    = c.
\end{align*}

Finally, we prove that $\lim_{\theta_{ii} \to 0} c_{ii}(\theta_{ii})=0$.
This holds because for any $\Theta_{-i,-i}$, the matrix with the $i$th row and column removed, $\Theta$ is positive-definite if and only if $\Theta_{-i,-i} \succ 0$ and $\theta_{ii} > \theta_{i,-i}^\top \Theta_{-i,-i}^{-1} \theta_{i,-i}$, where $\theta_{i,-i}$ is the $i$th column of $\Theta$ without the diagonal entry.

\subsection{Proof of Proposition \ref{prop:margindist}}

\begin{align*}
    \mbox{TV}(\pi_{ij},\pi^+_{ij}) &= \frac{1}{2} \int \abs{ \pi_{ij}(x) - \pi^+_{ij}(x) } \, dx 
    = \frac{1}{2} \int \pi_{ij}(x) \abs{ 1 - \frac{c_{ij}(x)}{c} } \, dx \\
    &= \frac{1}{2c} \int \pi_{ij}(x) \abs{ c_{ij(x)} - c } \, dx 
    = \frac{1}{2c} \mathbb{E}_{\pi_{ij}}\abs{ c_{ij}(X) - c}
\end{align*}
\begin{align*}
    KL(\,\pi_{ij}^+ \;||\; \pi_{ij}\,) &= \int \pi^+_{ij}(x) \log\left( \frac{\pi^+_{ij}(x)}{\pi_{ij}(x)} \right) \, dx \\
    &= \frac{1}{c} \int c_{ij}(x) \pi_{ij}(x) \log\left( \frac{c_{ij}(x)}{c} \right) \, dx 
    = \frac{1}{c} \mathbb{E}_{\pi_{ij}} \left[ c_{ij}(X) \log\left( \frac{c_{ij}(X)}{c} \right) \right]
\end{align*}

\subsection{Proof of Proposition \ref{prop:margindist_upperbound}}

We first prove the result for the general case, and subsequently for the identically-distributed case.
Recalling that $\pi^+_{ij}$ are the marginal densities of $p^+$, we write $\pi^+(\Theta) = \prod_{i \leq j} \pi^+_{ij}(\theta_{ij})$ as the density on $\Theta$ with independent entries and marginals $\pi^+_{ij}$. The KL divergence between $p^+$ and $p$ can then be written as
\begin{align*}
    KL(\, p^+ \;||\; p \,) &= 
    \int_{\calS} p^+(\Theta) \log\left( \frac{p^+(\Theta)}{p(\Theta)} \right) \, d\Theta \\
    &= \int_{\calS} p^+(\Theta) \log\left( \frac{p^+(\Theta)}{\pi^+(\Theta)} \right) \, d\Theta + \int_{\calS} p^+(\Theta) \log\left( \frac{\pi^+(\Theta)}{p(\Theta)} \right) \, d\Theta
\end{align*}
The first term is equal to $KL(\, p^+ \;||\; \pi^+ \,) \geq 0$. The second term is
\begin{align*}
    & \int_{\calS} p^+(\Theta) \log\left( \frac{\prod_{i \leq j} \pi_{ij}^+(\theta_{ij})}{\prod_{i \leq j} \pi_{ij}(\theta_{ij})} \right) \, d\Theta 
    &= \sum_{i \leq j} \int_{\calS} p^+(\Theta) \log\left( \frac{\pi_{ij}^+(\theta_{ij})}{\pi_{ij}(\theta_{ij
    })} \right) \, d\Theta 
    = \sum_{i \leq j} \int \pi_{ij}^+(\theta_{ij}) \log\left( \frac{\pi_{ij}^+(\theta_{ij})}{\pi_{ij}(\theta_{ij
    })} \right) \, d\theta_{ij}
\end{align*}
Notice that the integral in the last term is $KL(\, \pi^+_{ij} \;||\; \pi_{ij} \,)$. Hence we have
\begin{align*}
    KL(\, p^+ \;||\; p \,) &= 
    KL(\, p^+ \;||\; \pi^+ \,) + \sum_{i \leq j} KL(\, \pi^+_{ij} \;||\; \pi_{ij} \,)
    \geq \sum_{i \leq j} KL(\, \pi^+_{ij} \;||\; \pi_{ij} \,)
\end{align*}
Since $KL(\, p^+ \;||\; p \,) \leq - \log c$ from Proposition \ref{prop:dist}, we have that
\begin{align}\label{eq:proof_margindist_upperbound1}
    \sum_{i \leq j} KL(\,\pi_{ij}^+ \;||\; \pi_{ij}\,) \leq -\log c
\end{align}

The upper bound for the sum of TV distances then follows from Pinsker's inequality and the Bretagnolle–Huber inequality. Pinsker's inequality says that $\mbox{TV}(\pi_{ij},\pi^+_{ij}) \leq \sqrt{ \frac{1}{2} KL(\,\pi_{ij}^+ \;||\; \pi_{ij}\,) }$. Therefore
\begin{align*}
    \sum_{i \leq j} \mbox{TV}(\pi_{ij},\pi^+_{ij})
    \leq \frac{1}{\sqrt{2}} \sum_{i \leq j} \sqrt{KL(\,\pi_{ij}^+ \;||\; \pi_{ij}\,) }
    \leq \frac{1}{\sqrt{2}}  \sqrt{\sum_{i \leq j} KL(\,\pi_{ij}^+ \;||\; \pi_{ij}\,)}
    \leq \frac{1}{2}\sqrt{-k(k+1)\log c}
\end{align*}
the second inequality following from the Cauchy-Schwarz inequality, and the last inequality from \eqref{eq:proof_margindist_upperbound1}.

The Bretagnolle–Huber inequality says that $\mbox{TV}(\pi_{ij},\pi^+_{ij}) \leq \sqrt{ 1 - \exp\left( - KL(\,\pi_{ij}^+ \;||\; \pi_{ij}\,) \right) }$ and therefore
\begin{align*}
    \sum_{i \leq j} \mbox{TV}(\pi_{ij},\pi^+_{ij}) &\leq \sum_{i \leq j} \sqrt{ 1 - \exp\left( - KL(\,\pi_{ij}^+ \;||\; \pi_{ij}\,) \right) } \\
    &\leq \frac{k(k+1)}{2}\sqrt{ 1 - \exp\left( - \frac{2}{k(k+1)}\sum_{i \leq j}KL(\,\pi_{ij}^+ \;||\; \pi_{ij}\,) \right) }
    \\
    &\leq \frac{k(k+1)}{2}\sqrt{ 1 - \exp\left( \frac{2}{k(k+1)}\log c \right) }
    = \frac{k(k+1)}{2} \sqrt{1 - c^{2/k(k+1)} }.
\end{align*}
The second inequality holds because $\sqrt{1-\exp(x)}$ is increasing and concave on $[0,\infty)$. The third inequality is from \eqref{eq:proof_margindist_upperbound1}.

In the case of identically distributed entries, that is $\pi_{ii}=\pi_D$ and $\pi_{ij}=\pi$, we have that
$$ \sum_{i \leq j} KL(\, \pi^+_{ij} \;||\; \pi_{ij} \,) = k \, KL(\,\pi_{D}^+ \;||\; \pi_{D}\,) + \frac{k(k-1)}{2} \, KL(\,\pi^+ \;||\; \pi\,) \leq - \log c.$$
Since KL divergences are non-negative, we have that
\begin{align*}
    KL(\,\pi_{D}^+ \;||\; \pi_{D}\,) \leq - \frac{1}{k}\log c && KL(\,\pi^+ \;||\; \pi\,) \leq -\frac{2}{k(k-1)}\log c,
\end{align*}
which proves the desired result for the KL divergence. The bounds on the TV distance can be obtained by applying each of Pinsker's inequality and the Bretagnolle–Huber inequality along with these bounds on the KL divergence.




\section{Section \ref{sec:contolc} proofs}

\subsection{Proof of Proposition \ref{prop:cbounds}}

For the lower bound we have 
\begin{align*}
    c &\geq \mathbb{P}_p\left( \bigcap_{i < j} \left\{ \abs{\theta_{ij}} < \theta_{\min} / (k-1) \right\} \right) 
    = \mathbb{E}\left[ \mathbb{P}_p\left( \bigcap_{i < j} \left\{ \abs{\theta_{ij}} < \theta_{\min} / (k-1) \right\} \mid \theta_{\min} \right) \right]
\end{align*}
where the expectation is with respect to $\theta_{\min}$. 
Using that $\theta_{ij} \sim N(0,\sigma^2)$ independently across $i<j$,
the inner conditional probability is
\begin{align*}
    \prod_{i<j} \mathbb{P}_p\left( \abs{\theta_{ij}} < \theta_{\min} / (k-1) \mid \theta_{\min} \right)
    = \left( 2 \Phi \left( \theta_{\min}/(k-1) \sigma \right) - 1 \right)^{k(k-1)/2},
\end{align*}
proving the lower bound.
Similarly, for the upper bound we have 
\begin{align*}
    c &\leq \mathbb{P}_p\left( \bigcap_{i<j} \left\{ \abs{\theta_{ij}} < \theta_{\max} \right\} \right) 
    = \mathbb{E}\left[ \mathbb{P}_p\left( \bigcap_{i<j} \left\{ \abs{\theta_{ij}} < \theta_{\max} \right\} \mid \theta_{\max} \right) \right]
    \end{align*}
and the inner conditional probability is
\begin{align*}
    \prod_{i<j} \mathbb{P}_p\left( \abs{\theta_{ij}} < \theta_{\max} \mid \theta_{\max} \right) 
    = \left( 2 \Phi \left( \theta_{\max}/\sigma \right) - 1 \right)^{k(k-1)/2}.
\end{align*}

\subsection{Proof of Proposition \ref{prop:sigma}}

Consider two sets of continuous distributions for $i<j$ in the same location scale family with zero mean $\pi_{ij}$ with scale $\sigma$ and $\pi'_{ij}$ with scale $\sigma' > \sigma$ along with some diagonal densities $\pi_{ii}$. Let $p$ be the separable distribution with diagonal densities $\pi_{ii}$ and off-diagonal densities $\pi_{ij}$ and $p'$ be the separable distribution with diagonal densities $\pi_{ii}$ and off-diagonal densities $\pi'_{ij}$. Let $\Theta \sim p$ and let $\Theta'$ be the matrix with diagonal entries equal to $\Theta$ and off-diagonal entries $(\sigma'/\sigma)\theta_{ij}$. By the definition of location scale family, $\Theta'$ has distribution $p'$. 

The first part of the proof is showing that $c$ is decreasing in $\sigma$.
To do so, we prove that $\Theta \not\succ 0$ implies that $\Theta' \not\succ 0$ and therefore $\mathbb{P}_p(\Theta \succ 0) > \mathbb{P}_{p'}(\Theta' \succ 0)$, proving that $c$ is decreasing in $\sigma$.
To prove that $\Theta \not\succ 0$ implies that $\Theta' \not\succ 0$, write $\Theta= D + A$ where $D=\mbox{diag}(\Theta)$ and $A$ has zero diagonal and off-diagonal $a_{ij}=\theta_{ij}$.
We then have that 
$$\Theta'= D + (\sigma'/\sigma) A = \Theta + (\sigma'/\sigma - 1) A.$$
Let $x^*= \arg\min_{\| x \|=1} x^T \Theta x$ be the eigenvector associated to
$\lambda_{\min}(\Theta)= (x^*)^T \Theta x^*$, the smallest eigenvalue of $\Theta$.
We have that
\begin{align*}
   \lambda_{\min}(\Theta') \leq (x^*)^T \Theta' x^*=
   (x^*)^T \Theta x^* + (\sigma'/\sigma-1) (x^*)^T A x^*.
\end{align*}
Since $(x^*)^T \Theta x^*$ by the assumption that $\Theta \not\succ 0$.
It hence suffices to show that $(x^*)^T A x^* < 0$. This holds because
$$
\lambda_{\min}(\Theta)= (x^*)^T D x^* + (x^*)^T A x^* < 0
\Rightarrow
(x^*)^T A x^* < - (x^*)^T D x^* < 0,
$$
since the entries in $D$ are positive with probability 1.
The claim that $c$ is a continuous function of $\sigma$ follows from noting that
$\Theta= D + \sigma \widetilde{\Theta}$ where $\tilde{\theta}_{ii}=0$ and $\tilde{\theta}_{ij}$ has variance 1, and hence
\begin{align*}
&\lim_{\sigma \to a^+} c= 
\lim_{\sigma \to a^+} P\left( \min_{\| x \|=1} x^T (D + \sigma \widetilde{\Theta}) x > 0 \right)
= \int \lim_{\sigma \to a^+} P\left( \min_{\| x \|=1} x^T (D + \sigma \widetilde{\Theta}) x > 0 \mid D, \widetilde{\Theta} \right) dP(D, \widetilde{\Theta})
\\
&=\int \lim_{\sigma \to a^+} \mathbb{I} \left( \min_{\| x \|=1} x^T (D + \sigma \widetilde{\Theta}) x > 0 \mid D, \widetilde{\Theta} \right) dP(D, \widetilde{\Theta})
=\int \mathbb{I} \left( \min_{\| x \|=1} x^T (D + a \widetilde{\Theta}) x > 0 \mid D, \widetilde{\Theta} \right) dP(D, \widetilde{\Theta})
\\
& = P\left( \min_{\| x \|=1} x^T (D + a \widetilde{\Theta}) x > 0 \right),
\end{align*}
where in the second equality we used the dominated convergence theorem,
and in the fourth equality that $\min_{\| x \|=1} x^T (D + \sigma \widetilde{\Theta}) x$ is a continuous function of $\sigma$ (since $x^T (D + a \widetilde{\Theta}) x$ is continuous in $\sigma$, and eigenvalues are continuous functions of the input matrix).
Proceeding similarly shows that $\lim_{\sigma \to a^-} c$ is equal to the same quantity, and hence that $c$ is continuous in $\sigma$.

The second part of the proof is showing that $\lim_{\sigma \to 0} c= 1$ and $\lim_{\sigma \to \infty} c=0$. The former result follows from noting that, as $\sigma \to 0$, $\Theta \mid \mbox{diag}(\Theta) \stackrel{P}{\longrightarrow} \mbox{diag}(\Theta)$ and by the continuous mapping theorem $\lim_{\sigma \to 0}P(\Theta \succ 0 \mid \mbox{diag}(\Theta)) = 1$. The dominated convergence theorem then gives that 
$$\lim_{\sigma \to 0} P(\Theta \succ 0)= \int \lim_{\sigma \to 0} P(\Theta \succ 0 \mid \mbox{diag}(\Theta)) dP(\mbox{diag}(\Theta))= 1.$$
To show $\lim_{\sigma \to \infty} c=0$, since $\pi_{ij}$ is continuous with mean 0 and scale $\sigma$, letting $\sigma \to \infty$ implies that $\mathbb{P}(\abs{\theta_{ij}} \leq a) \to 0$ for any constant $a$. Recall that $\abs{\theta_{ij}} < \sqrt{\theta_{ii}\theta_{jj}}$ 
is a necessary condition for positive definiteness and let $a= \sqrt{\theta_{ii} \theta_{jj}}$. We have that
\begin{align*}
    c &\leq \mathbb{P}_p\left( \abs{\theta_{ij}} \leq a \right) 
    = \int \mathbb{P}_p\left( \abs{\theta_{ij}} \leq a \mid a \right) \pi(a) da.
\end{align*}
Since $\pi(a)$ does not depend on $\sigma$ and $\mathbb{P}_p\left( \abs{\theta_{ij}} \leq a \mid a \right) \leq 1$, we may use the dominated convergence theorem to obtain
\begin{align*}
    \lim_{\sigma \to a} c \leq \int \lim_{\sigma \to 0} \mathbb{P}_p\left( \abs{\theta_{ij}} \leq a \mid a \right) \pi(a) da= 0.
\end{align*}

\subsection{Proof of Theorem \ref{thm:Wigner}}

We decompose $\Theta$ as $ \Theta = \mu I + \sigma X_k $
where $X_k$ has zero diagonal and i.i.d.\ off-diagonals with density $\pi$. Standard Wigner matrix theory shows that
$$ W_k = \frac{X_k}{\sqrt{k}} = \frac{\Theta - \mu I}{\sigma \sqrt{k}}$$
has eigenvalues converging to the semicircle distribution and, in particular, has minimum eigenvalue $\lambda_\min(W_k) \to -2$ with probability 1 as $k \to \infty$ \citep{bai:1988}.

Since $\lambda_\min(\Theta) = \sigma\sqrt{k}\,\lambda_\min(W_k) + \mu$, we have
\begin{align*}
    c &= \mathbb{P}_p(\lambda_\min(\Theta) > 0) 
    = \mathbb{P}_p\left(\lambda_\min(W_k) > \frac{-\mu}{\sigma\sqrt{k}}\right).
\end{align*}

To conclude the proof, we show that if one sets any $\sigma = \frac{\mu}{(2+\delta_k) \sqrt{k}}$ where $\delta_k$ satisfies $k^{-2/3}= o(\delta_k)$, then $\lim_{k \to \infty} c= 1$.
This follows from the Wigner matrix theory in \cite{lee2014necessary}, which shows that deviations of the smallest eigenvalue of $\widetilde{\Theta}$ from $-2$ are of order $k^{-2/3}$ in probability. It follows that we still have $\lim_{k \to \infty} c= 1$ if one sets any $\sigma = \frac{\mu}{(2+\delta_k) \sqrt{k}}$ such that $k^{-2/3}= o(\delta_k)$.

\subsection{Proof of Theorem \ref{thm:Wigner2}}

We decompose $\Theta$ as
$\Theta = D + \sigma X_k = D_k + \sigma \sqrt{k} W_k$,
where $D$ is the diagonal of $\Theta$. From Weyl's inequality we have
\begin{align*}
    \lambda_\min(\Theta) &\geq \lambda_\min(D) + \sigma_k \sqrt{k} \lambda_\min(W_k) 
    = \theta_{\min,k} + \sigma_k \sqrt{k} \lambda_\min(W_k)
\end{align*}
It follows that
\begin{align*}
    c &= \mathbb{P}_p(\lambda_\min(\Theta) > 0)
    \geq \mathbb{P}\left( \frac{\theta_{\min,k}}{\sigma\sqrt{k}} > -\lambda_\min(W_k) \right)
\end{align*}
Since $-\lambda_\min(W_k) \to 2$ almost surely as $k \to \infty$, the first result follows.

The second result follows by a similar argument, but instead using the other Weyl inequality
\begin{align*}
    \lambda_\min(\Theta) &\leq \lambda_\max(D) + \sigma \sqrt{k} \lambda_\min(W_k) 
    = \theta_{\max,k} + \sigma \sqrt{k} \lambda_\min(W_k),
\end{align*}
so that
\begin{align*}
    c &= 1 - \mathbb{P}(\lambda_\min(\Theta) \leq 0) 
    \leq 1 - \mathbb{P}\left( \frac{\theta_{\max,k}}{\sigma\sqrt{k}} \leq -\lambda_\min(W_k) \right) 
    = \mathbb{P}\left( \frac{\theta_{\max,k}}{\sigma\sqrt{k}} > -\lambda_\min(W_k) \right)
\end{align*}

\section{Section \ref{sec:sparse_matrices} proofs}

\subsection{Proof of Proposition \ref{prop:mean_eigenval_z}}

To prove that $\mathbb E[ \lambda_{\min}(\Theta) \mid Z=z ] \geq \mathbb E[ \lambda_{\min}(\Theta) \mid Z=z']$,
it suffices to consider the case where $z'=z$ except for one entry $(i',j')$ such that $z_{i'j'}=0$ and $z'_{i',j'}=1$. From this, any other $z'$ such that $z' \geq z$ entry-wise follows by induction.
To ease notation, let $l_z(\Theta)$ and $l_{z'}(\Theta)$ be random variables with distribution equal to the conditional distribution $\lambda_{\min}(\Theta) \mid Z=z$ and $\lambda_{\min}(\Theta) \mid Z=z'$ respectively.
The goal is to show that $\mathbb E [l_z(\Theta)] \geq \mathbb E[l_{z'}(\Theta)]$.
The proof strategy is to express $l_z(\Theta)$ and $l_{z'}(\Theta)$ as infimums over a (conditionally Gaussian) random process, and then use the Sudakov-Fernique inequality to show that the expected infimums are ordered.

More precisely, the smallest eigenvalue of $\Theta$ for a given $Z=z$ is
\begin{align}
 l_z(\Theta)= \inf_{u: u^T u =1} u^T \Theta u=
\inf_{u: u^T u =1} \sum_{i=1}^p u_{ii}^2 \theta_{ii} + 2 \sum_{i>j} z_{ij} u_i u_j \theta_{ij}=
\inf_{u: u^T u =1} X_z(u),
\nonumber
\end{align}
where $X_z(u)$ is a Gaussian process for $u$ on the unit circle. 
Define $X_{z'}(u)$ analogously.

Conditionally on $D_\theta= \mbox{diag}(\theta_{11},\ldots,\theta_{pp})$
and on $\sigma^2= \{ \sigma_{ij}^2 \}_{i>j}$,
$X_z(u)$ and $X_{z'}(u)$ are Gaussian processes with equal means
$\mathbb E [ X_z(u) \mid D_\theta, \sigma^2 ] = \mathbb E [ X_{z'}(u) \mid D_\theta, \sigma^2 ] = \sum_{i=1}^p u_{ii}^2 \theta_{ii}$.
Define the conditionally-centered process
\begin{align}
 \widetilde{X}_z(u)= X_z(u) - \mathbb E [ X_z(u) \mid D_\theta, \sigma^2 ]= 2 \sum_{i > j} z_{ij} u_i u_j \theta_{ij},
\nonumber
\end{align}
so that $\mathbb E [\widetilde{X}_z(u) \mid D_\theta, \sigma^2] = 0$ and,
using that $\mathbb E[\widetilde{X}_z(u) - \widetilde{X}_z(v) \mid D_\theta, \sigma^2]=0$, we obtain that its increments are
\begin{align}
\mathbb E[ (\widetilde{X}_z(u) - \widetilde{X}_z(v))^2 \mid D_\theta, \sigma^2]=
\mathbb V \left( \sum_{i > j} 2 z_{ij} (u_iu_j - v_i v_j) \theta_{ij} \mid D_\theta, \sigma^2 \right)=
\sum_{i > j} 4 z_{ij} (u_iu_j - v_i v_j)^2 \sigma_{ij}^2,
\nonumber
\end{align}
since the $\theta_{ij}$'s are independent and also independent from $D_\theta$.

Note that $X_{z'}(u)$ has the same conditional mean as $X_z(u)$,
and that the corresponding conditionally-centered process is 
$\widetilde{X}_{z'}(u)= \widetilde{X}_z(u) + 2 u_{i'} u_{j'} \theta_{i'j'}$.
Its increments are
\begin{align}
\mathbb E[ (\widetilde{X}_z(u) - \widetilde{X}_z(v))^2 \mid D_\theta, \sigma^2]=
4 \sum_{i > j} z_{ij} (u_iu_j - v_i v_j)^2 \sigma_{ij}^2
+ (u_{i'} u_{j'} - v_{i'} v_{j'})^2 \sigma_{i'j'}^2.
\nonumber
\end{align}

The Sudakov-Fernique inequality (Theorem 7.2.11, \cite{vershynin:2018}) gives that
\begin{align}
\mathbb E \left[ \inf_{u: u^Tu=1} \widetilde{X}_{z'}(u) \mid D_\theta, \sigma^2  \right]
\leq \mathbb E \left[ \inf_{u: u^Tu=1} \widetilde{X}_z(u) \mid D_\theta, \sigma^2  \right],
\nonumber
\end{align}
which implies that
$\mathbb E \left[ \inf_{u: u^Tu=1} X_{z'}(u) \mid D_\theta, \sigma^2  \right]
\leq \mathbb E \left[ \inf_{u: u^Tu=1} X_z(u) \mid D_\theta, \sigma^2 \right]$
and therefore that
$$
\mathbb E[l_{z'}(\Theta) ] = \mathbb E \left[ \inf_{u: u^Tu=1} X_{z'}(u)  \right]
\leq \mathbb E \left[ \inf_{u: u^Tu=1} X_z(u) \right] \leq \mathbb E[l_z(\Theta) ],
$$
as we wished to prove.

To conclude the proof, we show that $\mathbb E[\lambda_\min(\Theta)]$ is decreasing in each $\eta_{ij}$.
Let $\eta$ and $\eta'$ such that $\eta_{ij} \leq \eta_{ij}'$ for all $i > j$,
and denote by $\mathbb E[\lambda_\min(\Theta)]$ and $\mathbb E[\lambda_\min'(\Theta)]$ the corresponding expectations.
Let $z(w)$ and $z'(w)$ with $(i,j)$ entry $z_{ij}(w_{ij}) = \mathbb{I}(w_{ij} < \eta_{ij})$ and $z_{ij}'(w_{ij})= \mathbb{I}(w_{ij} < \eta_{ij}')$ respectively for $i>j$, where $w_{ij} \sim \mbox{Unif}(0,1)$ and $w= \{ w_{ij} \}$.
Then we have that $z_{ij}(w_{ij}) \leq z_{ij}'(w_{ij})$ for all $i < j$ and marginally $z_{ij} \sim \mbox{Bern}(\eta_{ij})$, $z_{ij}' \sim \mbox{Bern}(\eta_{ij}')$.
As we already proved, it holds that
\begin{align}
\mathbb E [ \lambda_\min(\Theta)  \mid Z = z(w)] \geq \mathbb E [ \lambda_\min(\Theta)  \mid Z = z'(w)]
\nonumber
\end{align}
for all such $(z(w), z'(w))$, and therefore $ \mathbb E [ \lambda_\min(\Theta)]=$
\begin{align}
 \int \mathbb E [ \lambda_\min(\Theta)  \mid Z = z(w)] \prod_{i>j} \mbox{Unif}(w_{ij}; 0,1) d w_{ij}
\geq \int \mathbb E [ \lambda_\min(\Theta)  \mid Z = z'(w)] \prod_{i>j} \mbox{Unif}(w_{ij}; 0,1) d w_{ij}
\nonumber
\end{align}
$= \mathbb E [ \lambda_\min'(\Theta)]$
as we wished to prove.

\subsection{Proof of Lemma \ref{lem:eigenvalue_concentration}}



We can write $\Theta = \sigma \widetilde{\Theta}$, where
$\tilde{\theta}_{ii}= a / \sigma$ and $\tilde{\theta}_{ij} \mid Z_{ij}=1 \sim N(0,1)$. Hence
\begin{align}
 \mathbb P \left( |\lambda_{\min} (\Theta) - \mathbb E[\lambda_{\min} (\Theta) \mid Z=z] | > t \sigma \mid Z=z \right)=
 \mathbb P \left( |\lambda_{\min} (\widetilde{\Theta}) - \mathbb E[\lambda_{\min}(\widetilde{\Theta}) \mid Z=z] | > t \mid Z=z \right)
\leq e^{-t^2/2},
\nonumber
\end{align}
where the inequality follows from noting that $\lambda_{\min}(\widetilde{\Theta})$ is a 1-Lipschitz function of standard Gaussian variables, and therefore it is sub-Gaussian with unit pseudo-variance (Theorem 5.6, \cite{boucheron:2013}), and then applying the standard sub-Gaussian tail inequality (Theorem 2.1, \cite{boucheron:2013}).


\subsection{Proof of Corollary \ref{cor:cbounds_sparse}}

The proof follows directly from that of Proposition \ref{prop:cbounds} by noting that $\abs{\theta_{ij}} < \theta_\min / (k-1)$ and $\abs{\theta_{ij}} < \theta_\max$ are guaranteed for $z_{ij}=0$, leaving only $\sum_{i < j} z_{ij}$ terms.

\subsection{Proof of Corollary \ref{cor:sigma_sparse}}

The proof follows directly from that of Proposition \ref{prop:sigma} by considering $A$ and $\widetilde{\Theta}$ with sparsity pattern $Z=z$.

\subsection{Proof of Theorem \ref{thm:sparse_wigner_gaussian}}

The proof relies on Theorem 4.1.1 of \cite{tropp:2015}, reproduced as 
Theorem \ref{thm:concentration_matrix_gaussian} below, specialized to the case where $B_l$ are symmetric matrices, as is the case for us.

\begin{theorem} (Theorem 4.1.1, \cite{tropp:2015})
Let $B_l$ for $l=1,\ldots,L$ be fixed symmetric real matrices of dimension $k \times k$,
and let $U_l \sim N(0,1)$ independently for $l=1,\ldots,L$. Let $W= \sum_{l=1}^L U_l B_l$.
Then
\begin{align}
& \mathbb E \|W\| \leq \sqrt{2 \nu \log(2k)}
\nonumber \\
& \mathbb P( \| W \| \geq t) \leq 2k e^{-\frac{t^2}{2 \nu}},
\nonumber
\end{align}
for all $t>0$, where $\nu$ is the so-called matrix variance statistic
\begin{align}
 \nu= \| \mathbb E W^2 \|= \| \sum_{l=1}^L B_l^2 \|.
\nonumber
\end{align}
\label{thm:concentration_matrix_gaussian}
\end{theorem}

We prove each part of Theorem \ref{thm:sparse_wigner_gaussian} separately. 
We denote the operator norm (maximum absolute eigenvalue) of a symmetric positive-definite matrix $A$ by $\| A \|$.

\subsubsection*{{\bf Part (i)}}
Let $\Theta = \mu I + W$, where $W$ has zero diagonal and is symmetric with $W_{ij}= z_{ij} \theta_{ij}$ for $i \neq j$, and $z=(z_{ij})$ is fixed.
Since $\lambda_{\min}(\Theta)= \mu + \lambda_{\min}(W)$, we have that $\mathbb P(\lambda_{\min}(\Theta) \leq 0 \mid Z=z)= \mathbb P(\lambda_{\min}(W) \leq -\mu) \leq \mathbb P(\| W \| \geq \mu)$.
To bound $\mathbb P(\|W\| \geq \mu)$, the strategy is to write $W$ as a Gaussian random matrix series,
and then use Theorem 4.1.1 of \cite{tropp:2015} (reproduced above as Theorem \ref{thm:concentration_matrix_gaussian})
to bound its spectral norm $\| W \|$. Specifically, let
\begin{align}
 W= \sum_{i > j} z_{ij} \theta_{ij} (E_{ij} + E_{ji})
\nonumber
\end{align}
where $E_{ij}$ is the $k \times k$ matrix with $(i,j)$ entry equal to 1 and all other entries equal to zero,
and note that $\mathbb E[W]= 0$.
To apply Theorem \ref{thm:concentration_matrix_gaussian}, we need to find
\begin{align}
 \nu= \| \mathbb E(W^2) \|
=  \left\| \mathbb \sum_{i>j} \mathbb E\left([z_{ij} \theta_{ij} (E_{ij} + E_{ji})]^2\right) \right\|
=  \left\| \mathbb \sum_{i>j} z_{ij} (E_{ii} + E_{jj}) \mathbb E (\theta_{ij}^2) \right\|.
\nonumber
\end{align}
where we used that $z_{ij} \theta_{ij} (E_{ij} + E_{ji})$ are independent, $z_{ij}^2=z_{ij}$,
that simple algebra shows that $(E_{ij} + E_{ji})^2=(E_{ii}+E_{jj})$,
and that for any set independent and zero-mean random matrices $A_1,\ldots,A_n$, it holds that
\begin{align}
 \mathbb E \left( \left[ \sum_{i=1}^n A_i \right]^2 \right)=
\mathbb E \left( \sum_{i=1}^n A_i^2 + \sum_{i > j} A_i A_j \right)=
\mathbb E \left( \sum_{i=1}^n A_i^2 \right).
\nonumber
\end{align}

We hence obtain that
\begin{align}
& \nu= \left\| \sigma^2 \sum_{i>j} z_{ij} (E_{ii} + E_{jj})  \right\|=
\sigma^2 \left\| D  \right\|= \sigma^2 d_z,
\nonumber
\end{align}
where $D$ is a diagonal matrix with $(i,i)$ entry equal to $d_i= \sum_{i \neq j} z_{ij}$.
Hence, Theorem \ref{thm:concentration_matrix_gaussian} gives that
\begin{align}
&\mathbb P(l_z(\Theta) \leq 0) \leq \mathbb P( \| W \| \geq \mu) \leq 2 k e^{-0.5\mu^2/(\sigma^2 d_z)}.
\nonumber
\end{align}

\subsubsection*{{\bf Part (ii)}}
The result follows from Part (i) and the matrix Bernstein inequality (Theorem 1.6.2, \cite{tropp:2015}). 
More precisely, take $\Theta = \mu I + W$ and recall that in Part (i) we showed that
$W= \sum_{i>j} S_i$ where $S_i=\theta_{ij} (E_{ij} + E_{ji})$, where $E_{ij}$ and $E_{ji}$ have a single non-zero entry. 
From Gershgoryn's circle theorem, this implies that $\| S_i \| \leq a$, meeting the conditions of Theorem 1.6.2 in\cite{tropp:2015}. Taking the matrix variance statistic $\nu= \sigma^2 d_z$ obtained in Part (i), where $\sigma^2= V(\theta_{ij})$,
Theorem 1.6.2 in \cite{tropp:2015} gives
\begin{align}
 \mathbb P(\lambda_{\min}(\Theta) \leq 0 \mid Z=z)= 
 \mathbb P(\| W \| \geq \mu) \leq 2 k \exp \left\{ - \frac{\mu^2}{2 (\sigma^2 d_z + a \mu/3)} \right\}.
\nonumber
\end{align}

We proceed to prove that, if $\mu \geq 1$ and $\mu \geq 2 (\sigma^2 d_z + a/3) \log(2k/\alpha)$ then $c_z \geq 1-\alpha$.
To this end, we show that if these conditions on $\mu$ hold then we have that
$$
2 k \exp \left\{ - \frac{\mu^2}{2 (\sigma^2 d_z + a \mu/3)} \right\} \leq \alpha,
$$
and hence that $c_z \geq 1 - \alpha$. Equivalently, we show that
\begin{align*}
\frac{\mu^2}{2 (\sigma^2 d_z + a \mu/3)}  \geq \log\left(\frac{2k}{\alpha} \right).
\end{align*}
Since $\mu \geq 1$, the left-hand side above is $\geq 0.5 \mu^2/(\mu \sigma^2 d_z + a \mu/3)= 0.5 \mu /(\sigma^2 d_z + a/3)$. Therefore, it suffices that
$$
\frac{\mu}{2(\sigma^2 d_z + a/3)} \log\left(\frac{2k}{\alpha} \right).
$$

\subsubsection*{{\bf Part (iii)}}
Let $\Theta= D + W$ where $D$ is diagonal with $(i,i)$ entry equal to $\theta_{ii}$ and $W$ is as in Part (i).
Since $\lambda_{\min}(\Theta) \geq \lambda_{\min}(D) + \lambda_{\min}(W)= \theta_{\min} + \lambda_{\min}(W)$ where $\theta_{\min}= \min_{i=1,\ldots,k} \theta_{ii} \sim \mbox{Exp}(k / \mu)$, we have that
\begin{align}
& \mathbb P(\lambda_{\min}(\Theta) \leq 0 \mid Z=z) \leq P \left( \theta_{\min} + \lambda_{\min}(W) \leq 0 \right)
\nonumber \\
&= P \left( \lambda_{\min}(W) \leq -\theta_{\min} \mid \theta_{\min} \leq t \right) \mathbb P(\theta_{\min} \leq t)
+ P \left( \lambda_{\min}(W) \leq -\theta_{\min} \mid \theta_{\min} > t \right) \mathbb P(\theta_{\min} > t)
\nonumber \\
& \leq \mathbb P(\theta_{\min} \leq t)
+ P \left( \lambda_{\min}(W) \leq -t \right)=
1 - e^{- k t / \mu} + P \left( \lambda_{\min}(W) \leq -t \right),
\nonumber
\end{align}
for any $t \geq 0$. Using that $\mathbb P \left( \lambda_{\min}(W) \leq -t \right) \leq \mathbb P( \| W  \| \geq t ) $
and Theorem \ref{thm:concentration_matrix_gaussian} with $\nu= \sigma^2 d_z$ obtained in Part (i) gives that
\begin{align}
\mathbb P(\lambda_{\min}(\Theta) \leq 0 \mid Z=z) \leq 
1 - e^{-k t / \mu} + 2 k e^{-\frac{t^2}{2 \sigma^2 d_z}},
\label{eq:proof_sparse_wigner_tbound}
\end{align}
as we wished to prove.

Now, assume that $\sigma \sqrt{d_z} / \mu \leq 2\sqrt{2}$. Let
\begin{equation}
    t_1 = \sigma \sqrt{2 d_z \ln\left(\frac{2\sqrt{2} \mu}{ \sigma \sqrt{d_z}}\right)} \geq 0.
\nonumber
\end{equation}
Using that $1 - e^{-x} \leq x$ for $x \geq 0$ and plugging in $t=t_1$ gives, after simple algebra, that \eqref{eq:proof_sparse_wigner_tbound} is
\begin{equation}
\leq \frac{k \sigma}{\mu} \sqrt{2d_z \ln\left(\frac{2\sqrt{2} \mu}{ \sigma \sqrt{d_z}}\right)}
+  \frac{\sqrt{2} k \sigma \sqrt{d_z} }{\mu}
= \frac{\sqrt{2} k \sigma \sqrt{d_z}}{\mu} \left[ \sqrt{\ln\left(\frac{2\sqrt{2} \mu}{ \sigma \sqrt{d_z}}\right)} + 1 \right].
\nonumber
\end{equation}




To conclude the proof, we show that if $\mu \geq \frac{2 \sigma k \sqrt{d_z}}{\alpha} \sqrt{- W_{-1}\left( \frac{- \alpha^2}{32 e^2 k^2} \right)}$ then $c_z \geq 1 - \alpha$, where $W_{-1}()$ is the branch of Lambert's $W_q$ function associated to $q=-1$.
We shall use the bound on $c_z$ that we just obtained under the condition that $\mu \geq \sigma \sqrt{d_z}/(2\sqrt{2})$.
A useful fact is that $-W_{-1}(-1/e)=1$, and that $W_{-1}$ is decreasing on $(-1/e, 0)$, hence 
$$
\mu \geq \frac{2 \sigma k \sqrt{d_z}}{\alpha} \sqrt{- W_{-1}\left( \frac{- \alpha^2}{32 e^2 k^2} \right)} \geq \frac{2 \sigma k \sqrt{d_z}}{\alpha}
\geq \frac{\sigma \sqrt{d_z}}{2\sqrt{2}}.
$$

We first recall the definition of Lambert's W function. The equation $y e^y=x$ can be solved for $y$ only if $x \geq -1/e$. In that case $y= W_0(x)$ if $x \geq 0$, where $W_0$ is Lambert's $W_q$ function associated to $q=0$ (also called the principal branch). Further, if $-1/e \leq x < 0$, then there are two solutions, namely $y= W_0(x)$ and $y= W_{-1}(x)$.

Now, the goal is to show that for $\mu \geq \frac{2 \sigma k \sqrt{d_z}}{\alpha} \sqrt{- W_{-1}\left( \frac{- \alpha^2}{32 e^2 k^2} \right)}$ we have that
\begin{align}
\frac{k \sigma \sqrt{2 d_z}}{\mu} \left[ \sqrt{\ln\left(\frac{2\sqrt{2} \mu}{\sigma \sqrt{d_z}}\right)} + 1 \right] \leq \alpha,
\nonumber
\end{align}
which implies that $c_z \geq 1 - \alpha$.
To ease notation let $A=k \sigma \sqrt{2 d_z}$ and $B=\frac{2 \sqrt{2}}{\sigma \sqrt{d_z}}$.
Using that $\sqrt{\ln(B \mu)} + 1 \leq 2 \sqrt{\log(e B \mu)}$, 
it suffices to find $\mu$ such that
\begin{align}
\frac{A}{\mu} 2 \sqrt{\log \left(e B \mu \right)}
\leq \alpha 
\Leftrightarrow
\mu \geq \frac{2A}{\alpha} \sqrt{\log \left(e B \mu \right)}.
\nonumber
\end{align}
The rest of the proof simply rearranges the right-hand side to express it in terms of Lambert's $W_q$ function. To find the smallest $\mu$ satisfying this condition, we take
\begin{align}
\mu = \frac{2A}{\alpha} \sqrt{\log \left(e B \mu \right)}
\Leftrightarrow 
e^{\mu^2 \alpha^2/(2A)^2} = e B \mu
\Leftrightarrow
 \mu e^{-\frac{\mu^2 \alpha^2}{(2A)^2}}= \frac{1}{e B}
\Leftrightarrow
-\frac{\alpha^2}{2 A^2} \mu^2 e^{-\frac{\mu^2 \alpha^2}{2 A^2}}= -\frac{\alpha^2}{2 A^2} \frac{1}{e^2 B^2}.
 \nonumber
\end{align}
Let $y=- \alpha^2 \mu^2 / (2 A^2)$ and $x=-\alpha^2 / (2 e^2 A^2 B^2)$, the right-hand side above matches Lambert's $W_q$ function definition.
Note that $x < 0$, and a solution exists only if $x \geq -1/e$. Plugging in the expressions for $A$ and $B$, we have that
$
x =
- \frac{\alpha^2}{32 e^2 k^2}
$, which is $\geq -1/e$ since $k \geq 1$ and $\alpha \in [0,1]$.
Therefore, we may take the solution
\begin{align}
    y = W_{-1}(x)
    \Leftrightarrow
    \frac{- \alpha^2 \mu^2}{2 A^2} = W_{-1}\left(- \frac{\alpha^2}{32 e^2 k^2} \right)
    \Leftrightarrow
    \mu = \frac{\sqrt{2} A}{\alpha} \sqrt{-W_{-1}\left(- \frac{\alpha^2}{32 e^2  k^2} \right)},
    \nonumber
\end{align}
where $\sqrt{2} A= 2 k \sigma \sqrt{d_z}$, as we wished to prove.

\subsection{Proof of Corollary \ref{cor:tvd_sparse}}

By assumption we have $c_z \geq 1 - b(\mu,\sigma^2,d_z) \geq 1 - b(\mu,\sigma^2, \bar d)$, since $b$ is decreasing in $d_z$ by assumption. Hence
$$
c= \mathbb P_p(\Theta \succ 0)
= \sum_z \mathbb P_p(\Theta \succ 0 \mid Z=z) p(z)
= \sum_z c_z p(z)
\geq 1 - b(\mu, \sigma^2, \bar d).
$$

To prove the upper-bound on $KL(\,\pi^+ \;||\; \pi\,)$ and $\mbox{TV}(\pi, \pi^+)$, note that under the assumption that $p(z)=\prod_{i<j} \mbox{Bern}(\eta)$ we have a separable $p(\Theta)$ and PD-separable $p^+(\Theta)$. Hence, from Proposition \ref{prop:margindist_upperbound}, we have
\begin{align*}
&KL(\,\pi^+ \;||\; \pi\,) \leq -\frac{2 \log c}{k(k-1)}
\leq - \frac{2\log(1 - b(\mu,\sigma^2, \bar d))}{k(k-1)}
\\
&\mbox{TV}( \pi, \pi^+) \leq \min\left\{ \sqrt{\frac{-\log c}{k(k-1)}}, \sqrt{ 1 - e^{\frac{2 \log c}{k(k-1)}} } \right\} 
\leq \min\left\{ \sqrt{\frac{-\log(1 - b(\mu,\sigma^2, \bar d))}{k(k-1)}}, \sqrt{ 1 - e^{\frac{2 \log(1 - b(\mu,\sigma^2, \bar d))}{k(k-1)}} } \right\}
.
\end{align*}

\subsection{Proof of Corollary \ref{cor:tvd_sparse_suffcond}}

The results follow directly from Theorem \ref{thm:sparse_wigner_gaussian} and Corollary \ref{cor:tvd_sparse}.
Specifically the proof strategy is to note that, from Corollary \ref{cor:tvd_sparse}, $\mbox{TV}(p^+,p) \leq b(\mu,\sigma^2,\bar d)$
and
$$\mbox{TV}( \pi^+, \pi) 
\leq \sqrt{\frac{-\log(1 - b(\mu,\sigma^2, \bar d))}{k(k-1)}}$$
where $b(\mu,\sigma^2,\bar d)$ is given in Theorem \ref{thm:sparse_wigner_gaussian}. 
Denote $x = \mu / (\sigma \sqrt{\bar d})$.

\noindent {\bf Part (i).}
Under Theorem \ref{thm:sparse_wigner_gaussian}(i),
we have 
$ b(\mu,\sigma^2, \bar d) = 2ke^{-x^2/2}.$
Hence, if $\lim_{k \to \infty} x^2/2 - \log(k)= \infty$ then $\mbox{TV}^\infty(p,p^+) \leq \lim_{k \to \infty} b(\mu, \sigma^2, \bar d)=0$, as we wished to prove. 

For the marginal distributions, $\lim_{k \to \infty} \mu^2 / (2 \sigma^2 \bar d) - \log(k) > \log 2$ implies that $\lim_{k \to \infty} 1 - b(\mu,\sigma^2,\bar d) > 0$ and so there exists $\epsilon > 0$ such that eventually $1 - b(\mu,\sigma^2,\bar d) > \epsilon$. Hence,
\begin{align*}
\mbox{TV}(\pi^+, \pi) \leq \lim_{k \to \infty} \frac{-\log(1 - b(\mu,\sigma^2, \bar d))}{k(k-1)} \leq \lim_{k \to \infty} \frac{-\log(\epsilon)}{k(k-1)}=0.
\end{align*}


\noindent {\bf Part (ii).}
Under Theorem \ref{thm:sparse_wigner_gaussian}(ii),
we have 
$$
b(\mu, \sigma, \bar d)= 2 k \exp \left\{ - \frac{\mu^2}{2 (\sigma^2 \bar d + a \mu/3)} \right\}=
2 k \exp \left\{ - \frac{\mu^2}{2 \sigma^2 \bar d}  \frac{\sigma^2 \bar d}{(\sigma^2 \bar d + a \mu/3)} \right\}
\leq 2 k \exp \left\{ - \frac{(1 - \epsilon) \mu^2}{2 \sigma^2 \bar d} \right\}
$$
for any $\epsilon >0$ such that $1 - \epsilon \leq \frac{\sigma^2 \bar d}{(\sigma^2 \bar d + a \mu/3)}$, that is $\epsilon \geq \frac{a\mu/3}{\sigma^2 \bar d + a\mu/3}$. Since $\frac{a\mu/3}{\sigma^2 \bar d + a\mu/3}=o(1)$ by assumption, we may take arbitrarily small $\epsilon>0$ as $k \to \infty$.
The result then follows immediately from Part (i), replacing $\mu^2$ by $(1-\epsilon)\mu^2$.

\noindent {\bf Part (iii).}
Under Theorem \ref{thm:sparse_wigner_gaussian}(iii),
we have 
$ b(\mu,\sigma^2, \bar d) = \frac{\sqrt{2} k}{x} \left[ \sqrt{\log \left( 2\sqrt{2} x \right)}  + 1  \right].$
Since $x= ks_k$ where $s_k$ satisfies $\sqrt{\log k}= o(s_k)$, we have
$$
\mbox{TV}^\infty(p^+, p) \leq \lim_{k \to \infty} b(\mu, \sigma^2, \bar d)=
\lim_{k \to \infty} \frac{\sqrt{2}}{s_k}  \left[ \sqrt{\log \left( 2\sqrt{2} s_k k \right)}  + 1  \right] = 
\lim_{k \to \infty} \sqrt{2} {\frac{\sqrt{\log k}}{s_k}}
=0.
$$

Regarding $\mbox{TV}(\pi^+,\pi)$, since $x= \sqrt{2} k s_k$ where $s_k\geq \sqrt{\log k} + 2$, we have that
\begin{align*}
    b(\mu,\sigma^2,\bar d)= \frac{1}{s_k} \left[ \sqrt{\log(4 k s_k)}  + 1 \right].
\end{align*}
It is easy to check that, if one takes $s_k= \sqrt{\log k} + 2$, then $b(\mu,\sigma^2,\bar d) \leq 1$ for all $k \geq 2$.
Since $b(\mu,\sigma^2,\bar d)$ is decreasing in $s_k$, it follows that 
$b(\mu,\sigma^2,\bar d) \leq 1$ for any $s_k \geq \sqrt{\log k} + 2$.
Then $\log (1 - b(\mu,\sigma^2,\bar d))$ is well-defined and, using that $-\log(1 - x) \leq 1/\sqrt{1-x}$ for all $x \in [0,1]$, we have
\begin{align*}
    &\lim_{k \to \infty} \frac{-\log(1 - b(\mu, \sigma^2, \bar d))}{k(k-1)}
    \leq \lim_{k \to \infty} \frac{1}{k (k-1) \sqrt{1 - b(\mu, \sigma^2, \bar d)}}
    \nonumber \\
    &\leq
    \lim_{k \to \infty} \frac{(\sqrt{\log k} + 2)^{1/2}}{k (k-1) \sqrt{\sqrt{\log k} + 2 - \sqrt{\log(4 k [\sqrt{\log k} + 2]) - 1}}} 
     = 0,
\end{align*}
where in the last inequality we used that $b(\mu,\sigma^2,\bar d)$ is decreasing in $s_k$, and therefore the expression inside the limit is upper-bounded by evaluating $b(\mu, \sigma^2, \bar d)$ at $s_k= \sqrt{\log k} + 2$.



\subsection{Proof of Corollaries \ref{cor:SparseWigner}-\ref{cor:SparseWigner_randomdiag}}

\cite{erdHos2012spectral}, Theorem 2.7 shows that deviations of the maximum eigenvalue of a sparse Wigner matrix from $2$ are of the order $k^{-2/3}$. In particular, the maximum eigenvalue converges to $2$ as $k \to \infty$. Note the condition that $q > N^{1/3}$ where $q = \sqrt{k\eta_k}$ which corresponds to the condition $k^{-1/3} = o(\eta_k)$. The proof then follows directly by the same arguments as those of Theorems \ref{thm:Wigner}-\ref{thm:Wigner2}.

\end{document}